\documentclass[aps, preprint, superscriptaddress, journal=prl]{revtex4-2}
\usepackage{graphicx}
\usepackage{makecell, multirow}
\usepackage{color, bm, soul, xcolor}
\usepackage{amsmath, amssymb}
\usepackage{gensymb}
\usepackage{geometry}
\usepackage{braket}
\usepackage[colorlinks=true, linkcolor=black, citecolor=black, urlcolor=blue]{hyperref}
\newcommand{\ainse}{$\alpha$-In$_2$Se$_3$}
\newcommand{\bpinse}{$\beta'$-In$_2$Se$_3$}
\newcommand{\etal}{{\em et al}.\ }
 
\begin{document}

\title{Two-dimensional ferroelectric crystal with temperature-invariant ultralow thermal conductivity}

\author{Wenjie Zhou}
\affiliation{Zhejiang University, Hangzhou, Zhejiang 310058, China}
\affiliation{Department of Physics, School of Science, Westlake University, Hangzhou, Zhejiang 310030, China}

\author{Shi Liu}
\email{liushi@westlake.edu.cn}
\affiliation{Department of Physics, School of Science, Westlake University, Hangzhou, Zhejiang 310030, China}
\affiliation{Institute of Natural Sciences, Westlake Institute for Advanced Study, Hangzhou, Zhejiang 310024, China}

\begin{abstract}
We report the discovery of temperature-invariant ultralow thermal conductivity ($\kappa$) in monolayer $\beta'$-In$_2$Se$_3$, a two-dimensional ferroelectric crystal with in-plane polarization. Using a combination of generalized Wigner transport equation theory and machine-learning-assisted molecular dynamics simulations, we reveal that the balance between particle-like phonon propagating and wave-like tunneling transport mechanisms results in a propagating-tunneling-invariant (PTI) ultralow thermal conductivity of approximately 0.6 W/mK (comparable to that of glass) over a broad temperature range ($150<T<800$~K). This behavior stems from intrinsic strong lattice anharmonicity driven by ferroelectric dipolar fluctuations, eliminating the need for extrinsic structural modifications. In contrast, the \ainse~monolayer, which shares the same stoichiometry, exhibits a conventional temperature-dependent thermal conductivity, $\kappa (T) \propto T^{-1}$, typical of simple crystals. Furthermore, we demonstrate that the anharmonicity in $\beta'$-In$_2$Se$_3$~can be precisely modulated by an external electric field, enabling on-demand control of thermal transport properties, including modifying the temperature scaling behavior of heat conductivity and achieving a large thermal switching ratio of $\approx$2.5. These findings provide fundamental insights into the interplay between field-tunable lattice anharmonicity, phonon dynamics, and thermal transport mechanisms.
\end{abstract}

\maketitle
\newpage
Thermal conductivity ($\kappa$), a fundamental property of materials, is crucial in both scientific research and technological applications as it quantifies the efficiency of thermal energy transfer. Investigating its temperature ($T$) dependence helps elucidate transport mechanisms that differentiate metals, semiconductors, and insulators, as well as detect phase transitions~\cite{Lee17p371}. Effective thermal management is essential for industrial energy use~\cite{Dai21p2003734}, improving electronic device performance~\cite{He21p119223, Qian21p1188}, and advancing renewable energy technologies~\cite{Yan21p503, Jiang21p830}.

The theoretical framework for understanding thermal conductivity has evolved from Peierls's Boltzmann transport equation (BTE)~\cite{Peierls29p1055} to the Allen-Feldman (AF) model~\cite{Allen89p645} and, more recently, to the generalized Wigner transport equation (WTE)~\cite{Simoncelli19p809}. Grounded primarily in quasi-harmonic approximations of lattice vibrations, the BTE framework describes heat conduction in ordered crystals by treating phonons as particle-like entities undergoing diffuse and scattering affected by anharmonic interactions~\cite{Peierls29p1055}. This approach often predicts a $\kappa(T)\propto T^{-1}$ relationship when three-phonon processess dominate thermal resistance.
In contrast, the AF model is a harmonic theory of thermal transport in glasses where heat diffuses in a wave-like fashion through a Zener-like tunneling between quasi-degenerate vibrational eigenstates~\cite{Allen89p645}. In this model, the tunneling strength depends on the off-diagonal elements of the phonon velocity operator, whereas Peierls's BTE considers only diagonal elements. This wave-like transport mechanism explains the temperature-enhanced thermal conductivity and its eventual saturation in amorphous solids~\cite{Zhang22p96, Yang22p2200217}.
A recent breakthrough is the development of a generalized WTE framework, which unifies these approaches by simultaneously addressing  particle-like phonon propagating (population) and wave-like tunneling (coherence)~\cite{Simoncelli19p809}. This framework effectively captures both the effects of anharmonicity and disorder, providing a more holistic understanding of heat transport in crystalline materials with structural disorder.

The relative strength of propagating and tunneling transport mechanisms is determined by the interplay between phonon interband spacing, $
\Delta \omega = \left|\omega(\mathbf{q})_s - \omega(\mathbf{q})_{s'}\right|$, and linewidth difference, $\Delta \Gamma = \left|\Gamma(\mathbf{q})_s - \Gamma(\mathbf{q})_{s'}\right|$. Here, $\omega(\mathbf{q})_s$ represents the frequency of a vibrational mode $s$ at the wavevector $\mathbf{q}$, and $s\neq s'$. In simple crystalline materials where $\Delta \omega  \gg \frac{1}{2} \Delta \Gamma$, atomic vibrations are well described by energetically well-separated phonons. In this scenario, the particle-like propagating mechanism dominates thermal transport, leading to a thermal conductivity described by WTE that aligns with the BTE prediction. Conversely, in amorphous materials characterized by strong structural disorder where  $\Delta \omega  \ll \frac{1}{2} \Delta \Gamma$, the wave-like tunneling transport becomes the primary heat transfer mechanism. Recently, Simoncelli~\etal introduced the concept of ``propagating-tunneling-invariant" (PTI) thermal conductivity, identifying  materials where both mechanisms contribute nearly equally to thermal transport, resulting in temperature-independent thermal conductivity~\cite{Simoncelli24p1}. This discovery suggests a strategic approach to materials design: introducing an optimal degree of structural disorder in weakly anharmonic materials can balance particle-like and wave-like heat transport mechanisms, thereby maintaining stable thermal conductivity across a broad temperature range. 

In this work, we demonstrate the feasibility of achieving intrinsic PTI thermal conductivity in crystalline materials with simple chemical compositions, eliminating the need for fine-tuning structural disorder. Our model system is monolayer $\beta'$-In$_2$Se$_3$ featuring in-plane ferroelectricity~\cite{Ding17p14956, Zheng18peaar7720, Zhang21p2106951, Zhang19p8004}. We find that this two-dimensional (2D) ferroelectric displays pronounced temperature-driven order-disorder behavior at temperatures well below the Curie temperature ($T_c$). The thermally induced dipolar disorder generates strong intrinsic lattice anharmonicity, even in the absence of extrinsic structural defects. Importantly, the intrinsic strong anharmonicity activates the tunneling transport mechanism at an unexpectedly low temperature of $\approx$150 K. The balanced compensation between wave-like tunneling and particle-like propagating results in an ultralow in-plane thermal conductivity of $\kappa\approx0.6$ W/mK, which remains stable across a broad temperature range from 150 K to 800 K. In contrast, the ferroelectric $\alpha$-In$_2$Se$_3$ monolayer~\cite{Ding17p14956, Zhou17p5508, Xue18p1803738, Cui18p1253, Wan18p14885}, which shares the same chemical composition as the $\beta'$ phase, exhibits the conventional $\kappa(T)\propto T^{-1}$ behavior typical of simple crystals. Furthermore,  we show that the degree of anharmonicity in $\beta'$-In$_2$Se$_3$ can be tuned by applying an external electric field to suppress dipolar disorder, allowing for on-demand control over the relative strength of two transport mechanisms. Such robust field control not only achieves a giant thermal switch ratio of $\approx 2.5$ but also allows for the manipulation of the scaling law governing $\kappa(T)$ and $T$, a capability previously unattainable in conventional crystals or hybrid polymorphs.

We calculate the thermal conductivity with both the WTE theory and classical molecular dynamics (MD) simulations. The key quantities in the WTE framework are the second-order and third-order interatomic force constants (IFCs), which can be obtained through the finite difference method implemented in \texttt{Phono3py}~\cite{Togo15p094306, Togo23p353001, Chaput13p265506, Simoncelli19p809, Simoncelli22p041011}, utilizing atomic forces computed with density functional theory (DFT). However, the computational cost becomes demanding for crystals with low symmetry. To address this challenge, we develop a machine learning force field, known as neuroevolution potential (NEP)~\cite{Fan21p104309, Fan22p114801, Sonti24p8261}, for monolayer In$_2$Se$_3$. 
The NEP model features an input layer of descriptors representing local atomic environments, constructed from Chebyshev and Legendre polynomials. It includes a single hidden layer with 80 neurons that utilize a hyperbolic tangent activation function. The neural weights are optimized using the separable natural evolution strategy. The training database comprises 21955 monolayer configurations and 2808 bulk structures of various phases of In$_2$Se$_3$, with the energies and atomic forces computed using the projector augmented-wave formalism of DFT~\cite{Blochl94p17953} implemented in \texttt{VASP}~\cite{Kresse96p11169,Kresse96p15}.  We choose the PBE functional as the exchange-correlation functional~\cite{Perdew08p136406}, a kinetic energy cutoff of 700 eV, and a $k$-point grid density of 0.3~\AA$^{-1}$ (equivalent to a $4\times4\times1$ $k$-point mesh for a 20-atom slab model). 
The root mean square errors (RMSE) for training is 0.012 eV/atom for energy and 0.147 eV/\AA~for atomic forces. The NEP model demonstrates high accuracy in predicting a range of thermodynamic properties, phonon spectra, and temperature-driven phase transitions of In$_2$Se$_3$ polymorphs (see Supplementary Material).
Importantly, the temperature-dependent WTE thermal conductivity of the $\alpha$-In$_2$Se$_3$ monolayer, computed using NEP-derived IFCs, aligns well with values obtained from DFT-derived IFCs employing a $6\times6\times1$ supercell and a $60\times60\times1$ $q$-point grid. 
This agreement further validates the high accuracy of the NEP model. We estimate the conductivity via the homogeneous nonequilibrium molecular dynamics (HNEMD)~\cite{Fan19p064308} method based on the NEP model, which, in principle, captures all phonon-phonon interactions. All MD simulations are performed using the \texttt{GPUMD} package~\cite{Chen24p128, Fan22p114801} with a timestep of 1~fs. At each target temperature, the system is first equilibrated in the isothermal-isobaric ($NPT$) ensemble for at least 1 ns, followed by 10 ns of HNEMD runs in the canonical ($NVT$) ensemble 
employing a driving force parameter of $5\times10^{-5}$. 

\begin{figure}[htb]
\centering
\includegraphics[width=6.75 in]{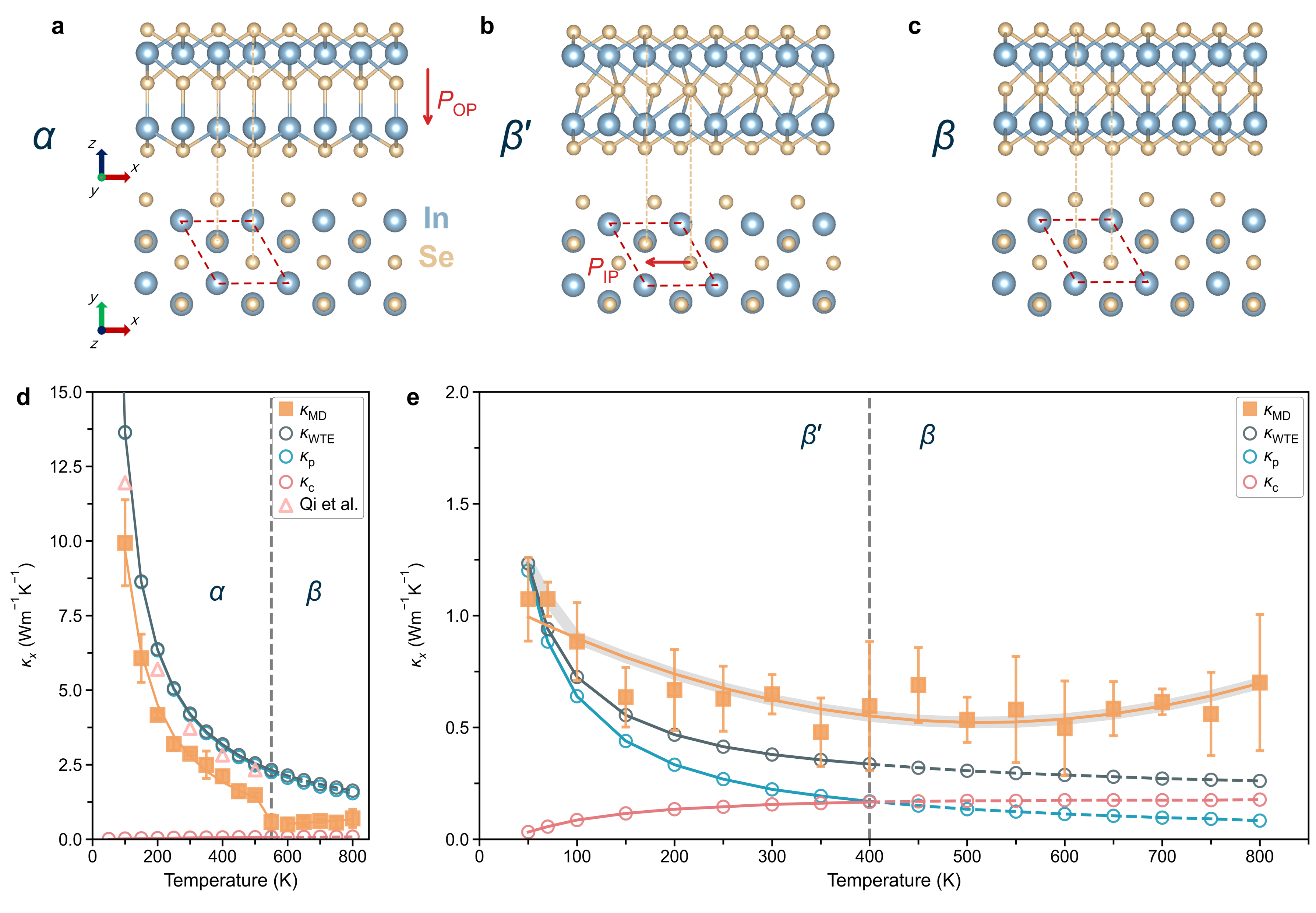}
 \caption{Temperature dependent thermal conductivity of monolayer In$_2$Se$_3$. Schematics of crystal structures of monolayer In$_2$Se$_3$ in the (a) $\alpha$, (b) $\beta^\prime$, and (c) $\beta$ phases. The red arrow represents the direction of local polarization. (d) Thermal conductivity of the $\alpha$ phase as a function of temperature, calculated using the WTE method ($\kappa_{\rm WTE}$, hollow circles) and the HNEMD method ($\kappa_{\rm MD}$, solid squares). 
 The $\kappa_{\rm WTE}$ is decomposed into the particle-like propagating (population) contribution ($\kappa_{\rm p}$) and wave-like tunneling (coherence) contribution ($\kappa_{\rm c}$). Since the WTE method does not capture the phase transition, the values of $\kappa_{\rm WTE}$ beyond $T_c\approx550$~K (vertical gray line) are connected by a dashed line. (e) Thermal conductivity of the $\beta'$ phase as a function of temperature. The vertical gray line marks the transition from the $\beta'$ phase to the $\beta$ phase at 400~K.}
  \label{fig_kappa}
\end{figure}

The model system featuring intrinsic PTI thermal conductivity that we identify is monolayer \bpinse.
Indium selenides represent a versatile class of materials that supports various stoichiometric ratios and has been extensively studied for thermoelectric applications~\cite{Nian21p033103, Han14p2747, Hung17p092107}. Recently, quasi-two-dimensional In$_2$Se$_3$ has garnered significant attention due to the room-temperature ferroic properties exhibited by its polymorphs, such as the $\alpha$ and $\beta'$ phases, even at the monolayer limit~\cite{Nian21p033103, Qi23p085701, Wu21p174107}. Specifically, monolayer In$_2$Se$_3$ consists of a covalently bonded quintuple layer with an atomic arrangement of Se-In-Se-In-Se. In the $\alpha$ phase, the central Se atoms are displaced out-of-plane (Fig.~\ref{fig_kappa}a), while in the $\beta'$ phase, they are displaced in-plane (Fig.~\ref{fig_kappa}b).
At elevated temperatures, both polar phases transition into the nonpolar $\beta$ phase (Fig.~\ref{fig_kappa}c).

We calculate the in-plane thermal conductivity ($\kappa_x$) along the $x$ direction for monolayer \ainse~and \bpinse~using both the WTE theory ($\kappa_{\rm WTE}$) and HNEMD ($\kappa_{\rm MD}$). As shown in Fig.~\ref{fig_kappa}d, for the $\alpha$ phase, both $\kappa_{\rm WTE}$ and $\kappa_{\rm MD}$ display a $T^{-1}$ dependence on temperature, characteristic of simple crystals. The values of $\kappa_{\rm WTE}$ computed using NEP-derived IFCs also agree closely with previously reported DFT values~\cite{Qi23p085701, Barbalinardo20p135104}. 
As expected, $\kappa_{\rm MD}$ is slightly lower than $\kappa_{\rm WTE}$, as MD simulations inherently account for all orders of phonon scattering processes. Our MD simulations capture the temperature-driven transition from the ferroelectric $\alpha$ phase to the paraelectric $\beta$ phase, which has been observed experimentally~\cite{Tao13p3501}. This transition is accompanied by a significant drop in $\kappa_{\rm MD}$ to 0.58 Wm$^{-1}$K$^{-1}$ beyond $T_c\approx 600$~K. In contrast, the WTE theory, which considers only thermal effects on phonon occupation numbers and dose not consider temperature-driven structural transitions, predicts a smooth decrease in $\kappa_{\rm WTE}$ with increasing temperature beyond $T_c$. To gain further insight, we decompose the total conductivity $\kappa_{\rm WTE}$ into contributions from phonon population ($\kappa_{\rm p}$) and coherence ($\kappa_{\rm c}$). This analysis reveals that $\kappa_{\rm WTE}$ is nearly identical to $\kappa_{\rm p}$, while $\kappa_{\rm c}$ is negligible, indicating the bonding interactions in \ainse~are highly harmonic with thermal transport dominated by particle-like phonon propagation.

Despite sharing the same chemical composition, the \bpinse~monolayer exhibits significantly lower thermal conductivity and a more complicated temperature dependence (Fig.~\ref{fig_kappa}e) compared to the $\alpha$ phase.
At $T = 50$~K, the values of $\kappa_{\rm WTE}$ and $\kappa_{\rm MD}$ show good agreement, both around $1.25$ Wm$^{-1}$K$^{-1}$. This is at least an order of magnitude lower than the thermal conductivity of the $\alpha$ phase, which is $33.7$ Wm$^{-1}$K$^{-1}$ at the same temperature. As the temperature rises, $\kappa_{\rm WTE}$ decays much more slowly than the typical $T^{-1}$ scaling. As illustrated in Fig.~\ref{fig_kappa}e, while the particle-like contribution $\kappa_{\rm p}$ adheres to the $T^{-1}$ dependence, the coherence contribution $\kappa_{\rm c}$ becomes increasingly significant for $T > 50$~K. For example, at $T = 100$~K, $\kappa_{\rm c}$ reaches 0.086 Wm$^{-1}$K$^{-1}$, accounting for approximately 12\% of the total thermal conductivity of 0.725 Wm$^{-1}$K$^{-1}$. The onset of a wave-like transport mechanism at such a low temperature is unusual and points to a strongly anharmonic lattice in the \bpinse~monolayer. Importantly, the compensating variations of $\kappa_{\rm p}$ and $\kappa_{\rm c}$ realizes the PTI thermal conductivity, where $\kappa_{\rm WTE}$ remains around 0.34 Wm$^{-1}$K$^{-1}$ with a standard deviation ($\sigma_{\kappa}$) of 0.08 Wm$^{-1}$K$^{-1}$ for temperatures above $150$~K.

Consistent with the results from WTE calcualtions, MD simulations also predict the emergence of temperature-invariant thermal conductivity in the \bpinse~monolayer. Remarkably, $\kappa_{\rm MD}$ remains nearly constant across the phase transition from the $\beta'$ phase to the nonpolar $\beta$ phase. This results in a wide temperature range, from 150 K to 800 K, where $\kappa_{\rm MD}$ stabilizes at approximately 0.60 Wm$^{-1}$K$^{-1}$ with $\sigma_{\kappa}=0.07$ Wm$^{-1}$K$^{-1}$.  Such temperature-invariant heat conductivity over a broad temperature range underscores the unique thermal transport properties of the \bpinse~monolayer. We observe that 
the values of $\kappa_{\rm{MD}}$ are consistently higher than those of $\kappa_{\rm{WTE}}$. This difference stems from the use of zero-Kelvin force constants in the WTE calculations to determine phonon frequencies and their temperature-dependent linewidths. Although the WTE approach includes anharmonic effects to some extent, it does not fully capture the stronger anharmonicity present in MD simulations, such as more pronounced negative frequency shifts and greater phonon broadening at elevated temperatures. Particularly, negative phonon frequency shifts tend to promote tunneling transport~\cite{Yang22p2200217}. As result, the WTE approach underestimates $\kappa_{\rm c}$ and, consequently, $\kappa_{\rm WTE}$, leading to systematically lower values compared to $\kappa_{\rm MD}$.

\begin{figure}[htb]
\centering
\includegraphics[width=6.75 in]{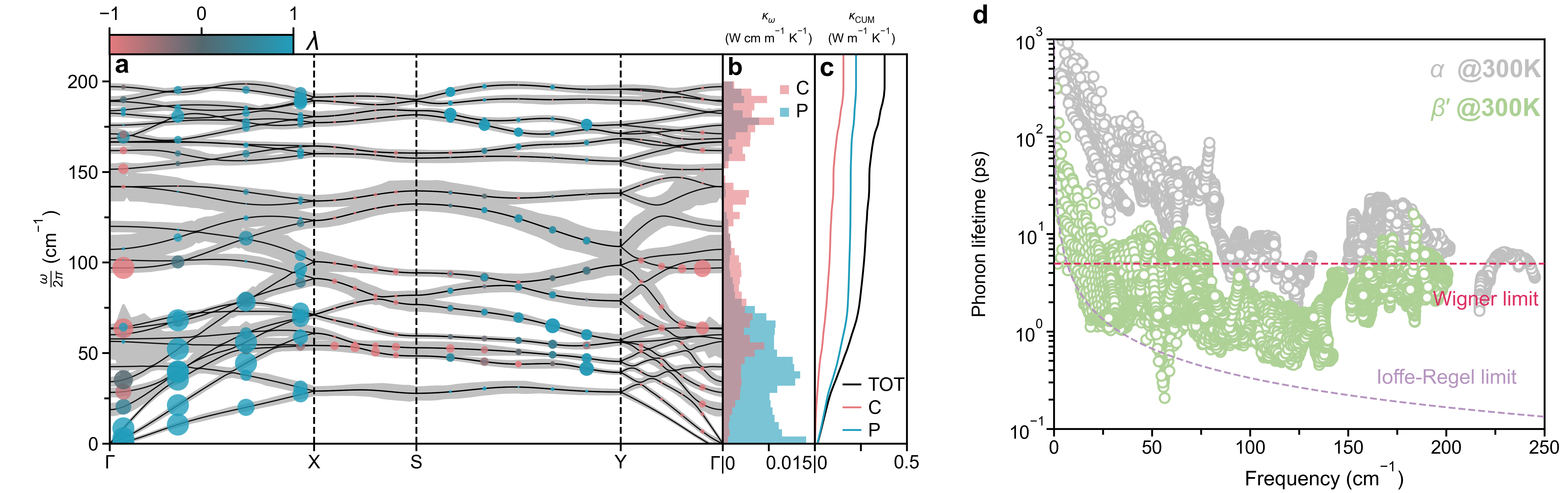}
 \caption{Phonon properties and heat conduction mechanisms in monolayer $\beta'$-In$_2$Se$_3$ at 300~K. (a) Phonon spectrum of the $\beta'$ phase with shared gray areas representing half the phonon linewidths ($\Gamma(\mathbf{q})_s$/2 for graphical clarity) at 300 K. The origin and contribution of thermal conductivity for representative phonon eigenstates are analyzed using the parameter, $\lambda = \frac{\kappa_{\rm p} - \kappa_{\rm c}}{\kappa_{\rm p} + \kappa_{\rm c}}$.
 The size of colored circles scales with the value of $\lambda$ and the color indicates the dominant contribution, blue for the particle-like population mechanism and red for the wave-like coherence mechanism. (b) Thermal conductivity density of states ($\kappa_{\omega}$) decomposed into contributions from the population mechanism (P, blue) and the coherence mechanism (C, red). (c) Cumulative total thermal conductivity ($\kappa_{\rm CUM}$, black) as the sum of the population contribution and coherence contribution. (d) Three-phonon lifetimes as a function of phonon frequency at 300 K, shown for the $\alpha$ phase (gray) and the $\beta'$ phase (green). The dashed red line represents the Wigner limit, while the dashed purple line marks the Ioffe-Regel limit.}
  \label{fig_wte-300K}
\end{figure}

The presence of PTI thermal conductivity in the \bpinse~monolayer is corroborated through an analysis of phonon properties at 300~K. The calculated phonon spectrum (Fig.~\ref{fig_wte-300K}a) with shaded gray areas representing half the phonon linewidth [$\Gamma(\mathbf{q})_s$/2] at 300~K , reveals low-energy flat branches with overlapping linewidths. To quantify the origin and contribution of thermal conductivity for a given phonon eigenstate $(\mathbf{q}, s)$, we define the parameter $\lambda = \frac{\kappa_{\rm p} - \kappa_{\rm c}}{\kappa_{\rm p} + \kappa_{\rm c}}$.
Phonons with large group velocities and low anharmonicity, characterized by small $\Gamma(\mathbf{q})_s$, predominantly contribute to the particle-like term ($\lambda\rightarrow 1$). In contrast, highly anharmonic flat bands with large $\Gamma(\mathbf{q})_s$ primarily contribute to the coherence term ($\lambda\rightarrow -1$). The thermal conductivity density of states (Fig.~\ref{fig_wte-300K}b) shows that $\kappa_{\rm p}$ is mainly dominated by low-frequency phonons below 75 cm$^{-1}$. However, the cumulative contributions from phonons across a broad frequency range make the coherence mechanism significant, accounting for approximately 40\% of the total conductivity (Fig.~\ref{fig_wte-300K}c). The role of phonon coherence is further elucidated by comparing the three-phonon lifetime ($\tau_{\rm 3ph}$) to two critical limits: the Wigner limit, given by $\tau_{\rm W} = 3N / \omega_{\rm max}$, and the Ioffe-Regel limit, defined as $\tau_{\rm IR} = \omega^{-1}$\cite{Camiola23p10, Das23p11521, Lucente23p033125, Pazhedath24p024064}. Here, $\omega_{\rm max}$ represents the maximum phonon frequency, and $N$ is the number of atoms in the primitive cell. As shown in Fig.~\ref{fig_wte-300K}d, most phonons lie above the Ioffe-Regel limit, confirming that the phonon picture remains valid and particle-like propagating still plays a significant role in thermal transport. Notably, a substantial number of phonons fall between the Wigner limit and the Ioffe-Regel limit, indicating that phonons are in a transitional regime,  neither fully particle-like nor entirely localized. This transitional behavior explains  the significant influence of phonon coherence on the thermal conductivity of \bpinse~at 300~K.

\begin{figure}[htb]
\centering
\includegraphics[width=6.75 in]{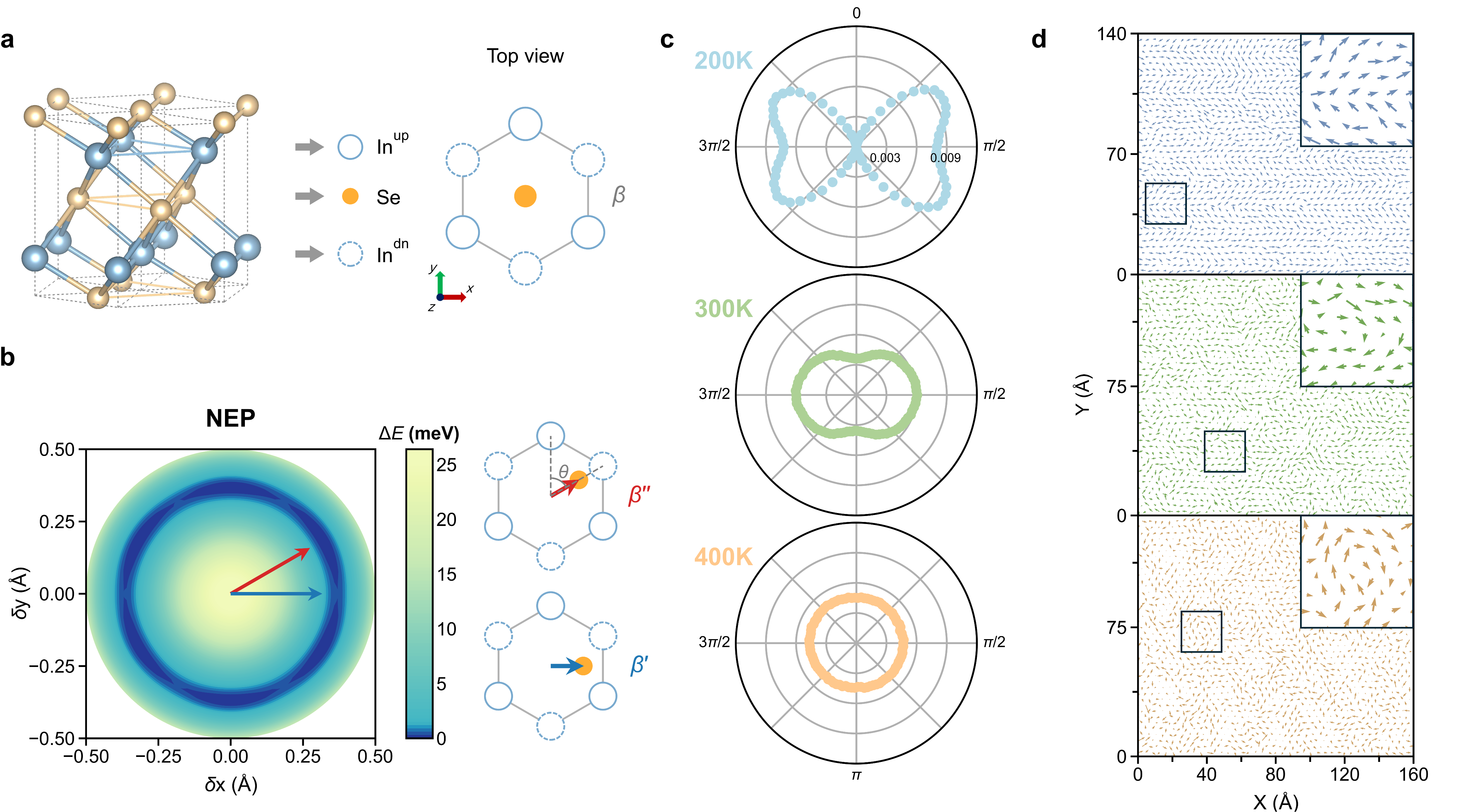}
 \caption{Dipolar disorder originated from a Mexican hat-shaped potential energy surface. (a) Schematic illustrations of side-view and top-view of the paraelectric $\beta$ phase. For clarity, the top view does not display the outermost Se atoms. (b) Calculated potential energy surface by displacing the central Se atom in the $\beta$-In$_2$Se$_3$ unit cell. The red arrow indicates the Se atom shift toward the hexagon corner, characteristic of the $\beta^{\prime\prime}$ phase, while the blue arrow represents the Se atom shift toward the hexagon edge, characteristic of the $\beta'$ phase. (c) Temperature dependence of $\Omega(\theta)\langle\rho(\theta)\rangle$, where $\Omega(\theta)$ is the probability density for Se displacements at angle $\theta$ and $\langle\rho(\theta)\rangle$ is the average magnitude of Se displacements for the same orientation. (d) Snapshots from MD simulations. Each arrow represents the local Se displacement in an 8000-atom supercell, with a zoomed-in view provided in the top-right corner.}
  \label{fig_polar}
\end{figure}

The onset of the PTI thermal conductivity at a relatively low temperature of $\approx$150~K suggests that monolayer \bpinse~displays strong lattice anharmonicity even under weak thermal fluctuations. We find that this anharmonicity arises from a Mexican hat-shaped potential energy surface (PES). Unit-cell DFT calculations further identify a  competing phase featuring in-plane polarization within the space group $C2$. We refer to this phase as $\beta''$ in this work. 
Both the $\beta'$ and $\beta''$ phases originate from the unstable phonon modes of the nonpolar $\beta$ phase (Fig.~\ref{fig_polar}a) at the center of the Brillouin zone~\cite{Ding17p14956, Zhang21p2106951, Liu19p025001}. The key distinction between these two phases lies in the direction of central-layer Se atom displacements when viewed from above (Fig.~\ref{fig_polar}b). In the $\beta'$ phase, the Se displacements point toward the midpoint between two neighboring In atoms, whereas in the $\beta''$ phase, the Se atoms are displaced directly toward one of the In atoms.
Notably, the $\beta''$ phase is only marginally higher in energy than the $\beta'$ phase by 1.2 meV/atom. 

The Mexican hat-shaped PES at finite temperatures menefest as strong dipolar disorder. 
We analyze the distributions of orientations ($\theta$) and magnitudes ($\rho$) of local Se displacements in the central layer, based on 1,000 instantaneous configurations sampled uniformly over a 100 ps MD simulation period for the equilibrated system. This approach provides a statistical perspective on the dynamic structure at a given temperature. To quantify the symmetry of the dynamic structure, we compute $\Omega(\theta)\left<\rho(\theta)\right>$, where $\Omega(\theta)$ represents the probability density of Se displacements with an in-plane orientation $\theta$, and $\left<\rho(\theta)\right>$ denotes the average magnitude of Se displacements in the same orientation. As shown in the polar plots of $\Omega(\theta)\left<\rho(\theta)\right>$ in Fig.~\ref{fig_polar}c, the butterfly-shaped distribution at 200~K reveals the presence of four variants of the $\beta''$ phase, corresponding to peaks at ${\pi}/{3}$, ${2\pi}/{3}$, ${4\pi}/{3}$, and ${5\pi}/{3}$, and two variants of the $\beta'$ phase, corresponding to peaks at ${\pi}/{2}$ and ${3\pi}/{2}$. Structurally, this distribution correspond to nanoscale antiferroelectric domains, as illustrated in the MD snapshot shown in the top panel of Fig.~\ref{fig_polar}d. These findings are consistent with previous experimental evidence of nanostriped antiferroelectric ordering in $\beta'$-In$_2$Se$_3$, as observed through real-space imaging and electric polarization mapping using scanning transmission electron microscopy~\cite{Xu20p047601, Zhang21p2106951}. The broad distribution of $\Omega(\theta)\left<\rho(\theta)\right>$ reflects the underlying Mexican hat-shaped energy landscape. 
At 300~K, the $\Omega(\theta)\left<\rho(\theta)\right>$ distribution becomes more isotropic, featuring two broad peaks at approximately ${\pi}/{2}$ and ${3\pi}/{2}$. Meanwhile, the peak value of $\left<\rho(\theta)\right>$ shifts to a lower magnitude compared to its value at 200 K, indicating weaker local polar ordering.  By 400 K, the in-plane orientation of Se displacements in the $\beta$ phase becomes fully isotropic, with nearly uniform population probabilities across all $\theta$ values. Nevertheless, the majority of the central-layer Se atoms remain locally displaced (bottom panel of Fig.~\ref{fig_polar}d), reflecting the order-disorder nature of the $\beta'$ to $\beta$ phase transition. It is noted that here the distribution of $\Omega(\theta)\left<\rho(\theta)\right>$ is employed as the order parameter to determine $T_c$.

\begin{figure}[htb]
\centering
\includegraphics[width=6.75 in]{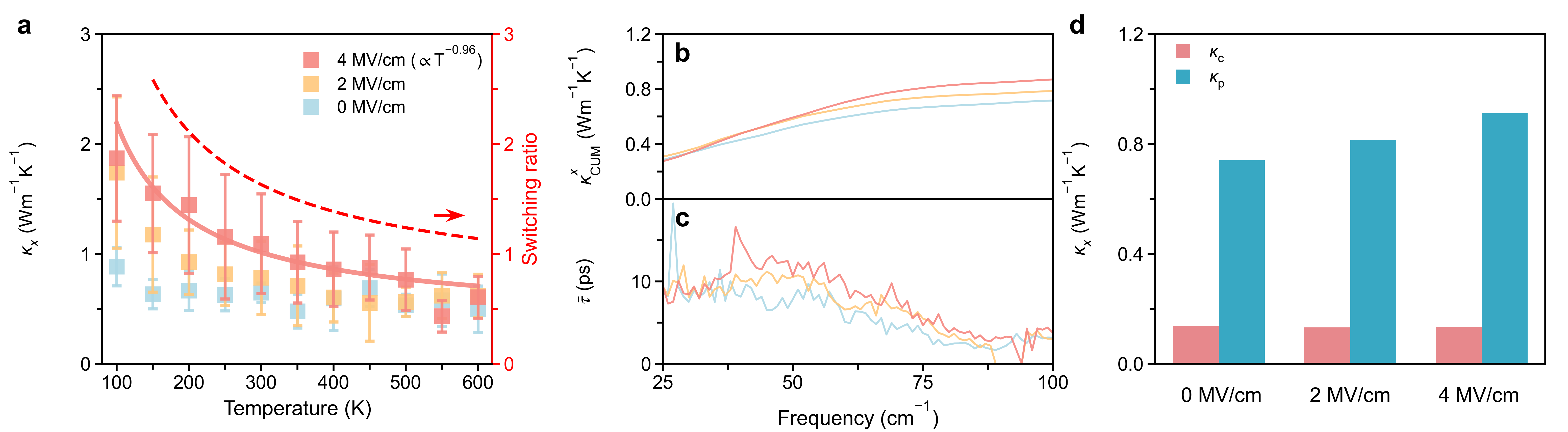}
 \caption{Electric field-tunable temperature dependence of in-plane thermal conductivity in the \bpinse~monolayer. 
 (a) Temperature dependence of $\kappa_x$ under electric fields applied along the $x$ axis. The red solid line represents a fit of $\kappa(T) \propto T^{-0.96}$. The red dashed line is the heat conductivity switching ratio at 4 MV/cm. 
(b) Cumulative thermal conductivity at 200 K for various electric field strengths, showing the contributions from different phonon frequencies.
(c) Frequency-resolved three-phonon lifetimes averaged over frequency bins of 5 cm$^{-1}$, considering only modes contributing more than 0.01\% to the total thermal conductivity.
(d)  Decomposition of thermal conductivity into population and coherence terms at 200~K under different electric field strengths.}
  \label{fig_AddElec}
\end{figure}

Finally, we demonstrate that the degree of anharmonicity and the relative contributions of the two heat transport mechanisms in \bpinse\ can be effectively modulated by applying an in-plane electric field. Figure~\ref{fig_AddElec}a shows the temperature dependence of $\kappa_x$ under an external electric field applied along the $x$-axis, for varying field strengths. 
The values of $\kappa_x$ are computed using equilibrium finite-field MD simulations combined the Kubo formula, incorporating the electric field via the using the ``force method"~\cite{Umari02p157602, Liu16p360}.
Our results reveal that $\kappa_x$ increases with the field strength at a fixed temperature. Notably, at 4 MV/cm, $\kappa_x$ exhibits a $T^{-1}$ dependence between 100 K and 600 K (see the solid line in Fig.~\ref{fig_AddElec}a), signaling a reemergence of particle-like propagating as the dominant heat conduction mechanism. 
As shown in the cumulative thermal conductivity (Fig.~\ref{fig_AddElec}b), the electric field mainly enhances the contributions from low-frequency phonons. Specifically, as computed with \texttt{Dynaphopy}~\cite{Carreras17p221, Zeng21p224307}, low-frequency phonons at 4 MV/cm exhibit longer three-phonon relaxation times compared to their zero-field values (Fig.~\ref{fig_AddElec}c), indicating that the electric field enhances lattice harmonicity. Additionally, we use the second-order IFCs derived from finite-field MD simulations within the generalized WTE framework to quantify the contributions of the two transport mechanisms. As presented in Fig.~\ref{fig_AddElec}d, the magnitude of $\kappa_{\rm p}$ increases with the field strength, while the contribution from the coherence mechanism remains largely unaffected. At 150 K, the heat conductivity switching ratio, defined as the ratio of $\kappa_x$ in the high-conductivity state to its value in the low-conductivity state, reaches a giant value of $\approx$2.5 (see the dashed line in Fig.~\ref{fig_AddElec}a), which we attribute to the field-induced modification of the temperature scaling behavior.

In summary, the intrinsic propagating-tunneling-invariant thermal conductivity observed in ferroelectric monolayer \bpinse~demonstrates that dipolar dynamics in two dimensions can be harnessed to engineer lattice anharmonicity and achieve unconventional thermal transport behavior in materials with simple chemical compositions.  The coexistence and compensation of particle-like and wave-like heat transport mechanisms, driven by the strong order-disorder behavior of dipoles arising from the Mexican-hat energy landscape, establish a design principle for realizing temperature-invariant thermal conductivity in crystalline systems, without relying on extrinsic structural modifications or defects.  This discovery challenges the conventional paradigms of thermal transport in solids by revealing how intrinsic anharmonic effects, induced by ferroelectric dipolar fluctuations, can dominate heat conduction even at relatively low temperatures.
The tunability of thermal transport via external electric fields further underscores the potential of \bpinse~for thermal management and novel device types such as thermal neurons~\cite{Nataf24p530}, where precise control of heat flow is critical. Beyond technological implications, it also serves a model system for exploring the fundamental physics of anharmonicity and phonon coherence in low-dimensional materials.

\begin{acknowledgments}
W.Z. and S.L. acknowledge the supports from Natural Science Foundation of China (52002335). The computational resource is provided by Westlake HPC Center. The authors sincerely thank Dr. Zekun Chen for his advice on the details of the thermal conductivity calculations.
\end{acknowledgments}

\clearpage
\newpage

\newpage

\newpage

\newpage

\newpage
\bibliography{SL}

\begin{thebibliography}{51}%
\makeatletter
\providecommand \@ifxundefined [1]{%
 \@ifx{#1\undefined}
}%
\providecommand \@ifnum [1]{%
 \ifnum #1\expandafter \@firstoftwo
 \else \expandafter \@secondoftwo
 \fi
}%
\providecommand \@ifx [1]{%
 \ifx #1\expandafter \@firstoftwo
 \else \expandafter \@secondoftwo
 \fi
}%
\providecommand \natexlab [1]{#1}%
\providecommand \enquote  [1]{``#1''}%
\providecommand \bibnamefont  [1]{#1}%
\providecommand \bibfnamefont [1]{#1}%
\providecommand \citenamefont [1]{#1}%
\providecommand \href@noop [0]{\@secondoftwo}%
\providecommand \href [0]{\begingroup \@sanitize@url \@href}%
\providecommand \@href[1]{\@@startlink{#1}\@@href}%
\providecommand \@@href[1]{\endgroup#1\@@endlink}%
\providecommand \@sanitize@url [0]{\catcode `\\12\catcode `\$12\catcode `\&12\catcode `\#12\catcode `\^12\catcode `\_12\catcode `\%12\relax}%
\providecommand \@@startlink[1]{}%
\providecommand \@@endlink[0]{}%
\providecommand \url  [0]{\begingroup\@sanitize@url \@url }%
\providecommand \@url [1]{\endgroup\@href {#1}{\urlprefix }}%
\providecommand \urlprefix  [0]{URL }%
\providecommand \Eprint [0]{\href }%
\providecommand \doibase [0]{https://doi.org/}%
\providecommand \selectlanguage [0]{\@gobble}%
\providecommand \bibinfo  [0]{\@secondoftwo}%
\providecommand \bibfield  [0]{\@secondoftwo}%
\providecommand \translation [1]{[#1]}%
\providecommand \BibitemOpen [0]{}%
\providecommand \bibitemStop [0]{}%
\providecommand \bibitemNoStop [0]{.\EOS\space}%
\providecommand \EOS [0]{\spacefactor3000\relax}%
\providecommand \BibitemShut  [1]{\csname bibitem#1\endcsname}%
\let\auto@bib@innerbib\@empty
\bibitem [{\citenamefont {Lee}\ \emph {et~al.}(2017)\citenamefont {Lee}, \citenamefont {Hippalgaonkar}, \citenamefont {Yang}, \citenamefont {Hong}, \citenamefont {Ko}, \citenamefont {Suh}, \citenamefont {Liu}, \citenamefont {Wang}, \citenamefont {Urban}, \citenamefont {Zhang}, \citenamefont {Dames}, \citenamefont {Hartnoll}, \citenamefont {Delaire},\ and\ \citenamefont {Wu}}]{Lee17p371}%
  \BibitemOpen
  \bibfield  {author} {\bibinfo {author} {\bibfnamefont {S.}~\bibnamefont {Lee}}, \bibinfo {author} {\bibfnamefont {K.}~\bibnamefont {Hippalgaonkar}}, \bibinfo {author} {\bibfnamefont {F.}~\bibnamefont {Yang}}, \bibinfo {author} {\bibfnamefont {J.}~\bibnamefont {Hong}}, \bibinfo {author} {\bibfnamefont {C.}~\bibnamefont {Ko}}, \bibinfo {author} {\bibfnamefont {J.}~\bibnamefont {Suh}}, \bibinfo {author} {\bibfnamefont {K.}~\bibnamefont {Liu}}, \bibinfo {author} {\bibfnamefont {K.}~\bibnamefont {Wang}}, \bibinfo {author} {\bibfnamefont {J.~J.}\ \bibnamefont {Urban}}, \bibinfo {author} {\bibfnamefont {X.}~\bibnamefont {Zhang}}, \bibinfo {author} {\bibfnamefont {C.}~\bibnamefont {Dames}}, \bibinfo {author} {\bibfnamefont {S.~A.}\ \bibnamefont {Hartnoll}}, \bibinfo {author} {\bibfnamefont {O.}~\bibnamefont {Delaire}},\ and\ \bibinfo {author} {\bibfnamefont {J.}~\bibnamefont {Wu}},\ }\bibfield  {title} {\bibinfo {title} {Anomalously low electronic thermal conductivity in metallic vanadium dioxide},\ }\href
  {https://doi.org/10.1126/science.aag0410} {\bibfield  {journal} {\bibinfo  {journal} {Science}\ }\textbf {\bibinfo {volume} {355}},\ \bibinfo {pages} {371–374} (\bibinfo {year} {2017})}\BibitemShut {NoStop}%
\bibitem [{\citenamefont {Dai}\ \emph {et~al.}(2021)\citenamefont {Dai}, \citenamefont {Lv}, \citenamefont {Ma}, \citenamefont {Wang}, \citenamefont {Ying}, \citenamefont {Yan}, \citenamefont {Tan}, \citenamefont {Gao}, \citenamefont {Xue}, \citenamefont {Yu}, \citenamefont {Yao}, \citenamefont {Wei}, \citenamefont {Sun}, \citenamefont {Wang}, \citenamefont {Liu}, \citenamefont {Chen}, \citenamefont {Xiang}, \citenamefont {Jiang}, \citenamefont {Xue}, \citenamefont {Wong}, \citenamefont {Maruyama},\ and\ \citenamefont {Lin}}]{Dai21p2003734}%
  \BibitemOpen
  \bibfield  {author} {\bibinfo {author} {\bibfnamefont {W.}~\bibnamefont {Dai}}, \bibinfo {author} {\bibfnamefont {L.}~\bibnamefont {Lv}}, \bibinfo {author} {\bibfnamefont {T.}~\bibnamefont {Ma}}, \bibinfo {author} {\bibfnamefont {X.}~\bibnamefont {Wang}}, \bibinfo {author} {\bibfnamefont {J.}~\bibnamefont {Ying}}, \bibinfo {author} {\bibfnamefont {Q.}~\bibnamefont {Yan}}, \bibinfo {author} {\bibfnamefont {X.}~\bibnamefont {Tan}}, \bibinfo {author} {\bibfnamefont {J.}~\bibnamefont {Gao}}, \bibinfo {author} {\bibfnamefont {C.}~\bibnamefont {Xue}}, \bibinfo {author} {\bibfnamefont {J.}~\bibnamefont {Yu}}, \bibinfo {author} {\bibfnamefont {Y.}~\bibnamefont {Yao}}, \bibinfo {author} {\bibfnamefont {Q.}~\bibnamefont {Wei}}, \bibinfo {author} {\bibfnamefont {R.}~\bibnamefont {Sun}}, \bibinfo {author} {\bibfnamefont {Y.}~\bibnamefont {Wang}}, \bibinfo {author} {\bibfnamefont {T.}~\bibnamefont {Liu}}, \bibinfo {author} {\bibfnamefont {T.}~\bibnamefont {Chen}}, \bibinfo {author} {\bibfnamefont {R.}~\bibnamefont
  {Xiang}}, \bibinfo {author} {\bibfnamefont {N.}~\bibnamefont {Jiang}}, \bibinfo {author} {\bibfnamefont {Q.}~\bibnamefont {Xue}}, \bibinfo {author} {\bibfnamefont {C.}~\bibnamefont {Wong}}, \bibinfo {author} {\bibfnamefont {S.}~\bibnamefont {Maruyama}},\ and\ \bibinfo {author} {\bibfnamefont {C.}~\bibnamefont {Lin}},\ }\bibfield  {title} {\bibinfo {title} {Multiscale structural modulation of anisotropic graphene framework for polymer composites achieving highly efficient thermal energy management},\ }\href {https://doi.org/10.1002/advs.202003734} {\bibfield  {journal} {\bibinfo  {journal} {Adv. Sci.}\ }\textbf {\bibinfo {volume} {8}},\ \bibinfo {pages} {2003734} (\bibinfo {year} {2021})}\BibitemShut {NoStop}%
\bibitem [{\citenamefont {He}\ \emph {et~al.}(2021)\citenamefont {He}, \citenamefont {Yan},\ and\ \citenamefont {Zhang}}]{He21p119223}%
  \BibitemOpen
  \bibfield  {author} {\bibinfo {author} {\bibfnamefont {Z.}~\bibnamefont {He}}, \bibinfo {author} {\bibfnamefont {Y.}~\bibnamefont {Yan}},\ and\ \bibinfo {author} {\bibfnamefont {Z.}~\bibnamefont {Zhang}},\ }\bibfield  {title} {\bibinfo {title} {Thermal management and temperature uniformity enhancement of electronic devices by micro heat sinks: A review},\ }\href {https://doi.org/10.1016/j.energy.2020.119223} {\bibfield  {journal} {\bibinfo  {journal} {Energy}\ }\textbf {\bibinfo {volume} {216}},\ \bibinfo {pages} {119223} (\bibinfo {year} {2021})}\BibitemShut {NoStop}%
\bibitem [{\citenamefont {Qian}\ \emph {et~al.}(2021)\citenamefont {Qian}, \citenamefont {Zhou},\ and\ \citenamefont {Chen}}]{Qian21p1188}%
  \BibitemOpen
  \bibfield  {author} {\bibinfo {author} {\bibfnamefont {X.}~\bibnamefont {Qian}}, \bibinfo {author} {\bibfnamefont {J.}~\bibnamefont {Zhou}},\ and\ \bibinfo {author} {\bibfnamefont {G.}~\bibnamefont {Chen}},\ }\bibfield  {title} {\bibinfo {title} {Phonon-engineered extreme thermal conductivity materials},\ }\href {https://doi.org/10.1038/s41563-021-00918-3} {\bibfield  {journal} {\bibinfo  {journal} {Nat. Mater.}\ }\textbf {\bibinfo {volume} {20}},\ \bibinfo {pages} {1188–1202} (\bibinfo {year} {2021})}\BibitemShut {NoStop}%
\bibitem [{\citenamefont {Yan}\ and\ \citenamefont {Kanatzidis}(2021)}]{Yan21p503}%
  \BibitemOpen
  \bibfield  {author} {\bibinfo {author} {\bibfnamefont {Q.}~\bibnamefont {Yan}}\ and\ \bibinfo {author} {\bibfnamefont {M.~G.}\ \bibnamefont {Kanatzidis}},\ }\bibfield  {title} {\bibinfo {title} {High-performance thermoelectrics and challenges for practical devices},\ }\href {https://doi.org/10.1038/s41563-021-01109-w} {\bibfield  {journal} {\bibinfo  {journal} {Nat. Mater.}\ }\textbf {\bibinfo {volume} {21}},\ \bibinfo {pages} {503–513} (\bibinfo {year} {2021})}\BibitemShut {NoStop}%
\bibitem [{\citenamefont {Jiang}\ \emph {et~al.}(2021)\citenamefont {Jiang}, \citenamefont {Yu}, \citenamefont {Cui}, \citenamefont {Liu}, \citenamefont {Xie}, \citenamefont {Liao}, \citenamefont {Zhang}, \citenamefont {Huang}, \citenamefont {Ning}, \citenamefont {Jia}, \citenamefont {Zhu}, \citenamefont {Bai}, \citenamefont {Chen}, \citenamefont {Pennycook},\ and\ \citenamefont {He}}]{Jiang21p830}%
  \BibitemOpen
  \bibfield  {author} {\bibinfo {author} {\bibfnamefont {B.}~\bibnamefont {Jiang}}, \bibinfo {author} {\bibfnamefont {Y.}~\bibnamefont {Yu}}, \bibinfo {author} {\bibfnamefont {J.}~\bibnamefont {Cui}}, \bibinfo {author} {\bibfnamefont {X.}~\bibnamefont {Liu}}, \bibinfo {author} {\bibfnamefont {L.}~\bibnamefont {Xie}}, \bibinfo {author} {\bibfnamefont {J.}~\bibnamefont {Liao}}, \bibinfo {author} {\bibfnamefont {Q.}~\bibnamefont {Zhang}}, \bibinfo {author} {\bibfnamefont {Y.}~\bibnamefont {Huang}}, \bibinfo {author} {\bibfnamefont {S.}~\bibnamefont {Ning}}, \bibinfo {author} {\bibfnamefont {B.}~\bibnamefont {Jia}}, \bibinfo {author} {\bibfnamefont {B.}~\bibnamefont {Zhu}}, \bibinfo {author} {\bibfnamefont {S.}~\bibnamefont {Bai}}, \bibinfo {author} {\bibfnamefont {L.}~\bibnamefont {Chen}}, \bibinfo {author} {\bibfnamefont {S.~J.}\ \bibnamefont {Pennycook}},\ and\ \bibinfo {author} {\bibfnamefont {J.}~\bibnamefont {He}},\ }\bibfield  {title} {\bibinfo {title} {High-entropy-stabilized chalcogenides with high
  thermoelectric performance},\ }\href {https://doi.org/10.1126/science.abe1292} {\bibfield  {journal} {\bibinfo  {journal} {Science}\ }\textbf {\bibinfo {volume} {371}},\ \bibinfo {pages} {830–834} (\bibinfo {year} {2021})}\BibitemShut {NoStop}%
\bibitem [{\citenamefont {Peierls}(1929)}]{Peierls29p1055}%
  \BibitemOpen
  \bibfield  {author} {\bibinfo {author} {\bibfnamefont {R.}~\bibnamefont {Peierls}},\ }\bibfield  {title} {\bibinfo {title} {Zur kinetischen theorie der wärmeleitung in kristallen},\ }\href {https://doi.org/10.1002/andp.19293950803} {\bibfield  {journal} {\bibinfo  {journal} {Ann. Phys.}\ }\textbf {\bibinfo {volume} {395}},\ \bibinfo {pages} {1055} (\bibinfo {year} {1929})}\BibitemShut {NoStop}%
\bibitem [{\citenamefont {Allen}\ and\ \citenamefont {Feldman}(1989)}]{Allen89p645}%
  \BibitemOpen
  \bibfield  {author} {\bibinfo {author} {\bibfnamefont {P.~B.}\ \bibnamefont {Allen}}\ and\ \bibinfo {author} {\bibfnamefont {J.~L.}\ \bibnamefont {Feldman}},\ }\bibfield  {title} {\bibinfo {title} {Thermal conductivity of glasses: Theory and application to amorphous si},\ }\href {https://doi.org/10.1103/PhysRevLett.62.645} {\bibfield  {journal} {\bibinfo  {journal} {Phys. Rev. Lett.}\ }\textbf {\bibinfo {volume} {62}},\ \bibinfo {pages} {645} (\bibinfo {year} {1989})}\BibitemShut {NoStop}%
\bibitem [{\citenamefont {Simoncelli}\ \emph {et~al.}(2019)\citenamefont {Simoncelli}, \citenamefont {Marzari},\ and\ \citenamefont {Mauri}}]{Simoncelli19p809}%
  \BibitemOpen
  \bibfield  {author} {\bibinfo {author} {\bibfnamefont {M.}~\bibnamefont {Simoncelli}}, \bibinfo {author} {\bibfnamefont {N.}~\bibnamefont {Marzari}},\ and\ \bibinfo {author} {\bibfnamefont {F.}~\bibnamefont {Mauri}},\ }\bibfield  {title} {\bibinfo {title} {Unified theory of thermal transport in crystals and glasses},\ }\href {https://doi.org/10.1038/s41567-019-0520-x} {\bibfield  {journal} {\bibinfo  {journal} {Nat. Phys.}\ }\textbf {\bibinfo {volume} {15}},\ \bibinfo {pages} {809} (\bibinfo {year} {2019})}\BibitemShut {NoStop}%
\bibitem [{\citenamefont {Zhang}\ \emph {et~al.}(2022)\citenamefont {Zhang}, \citenamefont {Guo}, \citenamefont {Bescond}, \citenamefont {Chen}, \citenamefont {Nomura},\ and\ \citenamefont {Volz}}]{Zhang22p96}%
  \BibitemOpen
  \bibfield  {author} {\bibinfo {author} {\bibfnamefont {Z.}~\bibnamefont {Zhang}}, \bibinfo {author} {\bibfnamefont {Y.}~\bibnamefont {Guo}}, \bibinfo {author} {\bibfnamefont {M.}~\bibnamefont {Bescond}}, \bibinfo {author} {\bibfnamefont {J.}~\bibnamefont {Chen}}, \bibinfo {author} {\bibfnamefont {M.}~\bibnamefont {Nomura}},\ and\ \bibinfo {author} {\bibfnamefont {S.}~\bibnamefont {Volz}},\ }\bibfield  {title} {\bibinfo {title} {How coherence is governing diffuson heat transfer in amorphous solids},\ }\href {https://doi.org/10.1038/s41524-022-00776-w} {\bibfield  {journal} {\bibinfo  {journal} {npj Comput. Mater.}\ }\textbf {\bibinfo {volume} {8}},\ \bibinfo {pages} {96} (\bibinfo {year} {2022})}\BibitemShut {NoStop}%
\bibitem [{\citenamefont {Yang}\ and\ \citenamefont {Cao}(2022)}]{Yang22p2200217}%
  \BibitemOpen
  \bibfield  {author} {\bibinfo {author} {\bibfnamefont {L.}~\bibnamefont {Yang}}\ and\ \bibinfo {author} {\bibfnamefont {B.-Y.}\ \bibnamefont {Cao}},\ }\bibfield  {title} {\bibinfo {title} {Significant anharmonicity of thermal transport in amorphous silica at high temperature},\ }\href {https://doi.org/https://doi.org/10.1002/pssr.202200217} {\bibfield  {journal} {\bibinfo  {journal} {Phys. Status Solidi RRL}\ }\textbf {\bibinfo {volume} {16}},\ \bibinfo {pages} {2200217} (\bibinfo {year} {2022})}\BibitemShut {NoStop}%
\bibitem [{\citenamefont {Simoncelli}\ \emph {et~al.}(2024)\citenamefont {Simoncelli}, \citenamefont {Fournier}, \citenamefont {Marangolo} \emph {et~al.}}]{Simoncelli24p1}%
  \BibitemOpen
  \bibfield  {author} {\bibinfo {author} {\bibfnamefont {M.}~\bibnamefont {Simoncelli}}, \bibinfo {author} {\bibfnamefont {D.}~\bibnamefont {Fournier}}, \bibinfo {author} {\bibfnamefont {M.}~\bibnamefont {Marangolo}}, \emph {et~al.},\ }\bibfield  {title} {\bibinfo {title} {Temperature-invariant heat conductivity from compensating crystalline and glassy transport: from the steinbach meteorite to furnace bricks},\ }\bibfield  {journal} {\bibinfo  {journal} {Preprint at Research Square}\ }\href {https://doi.org/10.21203/rs.3.rs-4456620/v1} {10.21203/rs.3.rs-4456620/v1} (\bibinfo {year} {2024})\BibitemShut {NoStop}%
\bibitem [{\citenamefont {Ding}\ \emph {et~al.}(2017)\citenamefont {Ding}, \citenamefont {Zhu}, \citenamefont {Wang}, \citenamefont {Gao}, \citenamefont {Xiao}, \citenamefont {Gu}, \citenamefont {Zhang},\ and\ \citenamefont {Zhu}}]{Ding17p14956}%
  \BibitemOpen
  \bibfield  {author} {\bibinfo {author} {\bibfnamefont {W.}~\bibnamefont {Ding}}, \bibinfo {author} {\bibfnamefont {J.}~\bibnamefont {Zhu}}, \bibinfo {author} {\bibfnamefont {Z.}~\bibnamefont {Wang}}, \bibinfo {author} {\bibfnamefont {Y.}~\bibnamefont {Gao}}, \bibinfo {author} {\bibfnamefont {D.}~\bibnamefont {Xiao}}, \bibinfo {author} {\bibfnamefont {Y.}~\bibnamefont {Gu}}, \bibinfo {author} {\bibfnamefont {Z.}~\bibnamefont {Zhang}},\ and\ \bibinfo {author} {\bibfnamefont {W.}~\bibnamefont {Zhu}},\ }\bibfield  {title} {\bibinfo {title} {Prediction of intrinsic two-dimensional ferroelectrics in {In$_2$Se$_3$} and other {III$_2$-VI$_3$} van der waals materials},\ }\href {https://doi.org/10.1038/ncomms14956} {\bibfield  {journal} {\bibinfo  {journal} {Nat. Commun.}\ }\textbf {\bibinfo {volume} {8}},\ \bibinfo {pages} {14956} (\bibinfo {year} {2017})}\BibitemShut {NoStop}%
\bibitem [{\citenamefont {Zheng}\ \emph {et~al.}(2018)\citenamefont {Zheng}, \citenamefont {Yu}, \citenamefont {Zhu}, \citenamefont {Collins}, \citenamefont {Kim}, \citenamefont {Lou}, \citenamefont {Xu}, \citenamefont {Li}, \citenamefont {Wei}, \citenamefont {Zhang}, \citenamefont {Edmonds}, \citenamefont {Li}, \citenamefont {Seidel}, \citenamefont {Zhu}, \citenamefont {Liu}, \citenamefont {Tang},\ and\ \citenamefont {Fuhrer}}]{Zheng18peaar7720}%
  \BibitemOpen
  \bibfield  {author} {\bibinfo {author} {\bibfnamefont {C.}~\bibnamefont {Zheng}}, \bibinfo {author} {\bibfnamefont {L.}~\bibnamefont {Yu}}, \bibinfo {author} {\bibfnamefont {L.}~\bibnamefont {Zhu}}, \bibinfo {author} {\bibfnamefont {J.~L.}\ \bibnamefont {Collins}}, \bibinfo {author} {\bibfnamefont {D.}~\bibnamefont {Kim}}, \bibinfo {author} {\bibfnamefont {Y.}~\bibnamefont {Lou}}, \bibinfo {author} {\bibfnamefont {C.}~\bibnamefont {Xu}}, \bibinfo {author} {\bibfnamefont {M.}~\bibnamefont {Li}}, \bibinfo {author} {\bibfnamefont {Z.}~\bibnamefont {Wei}}, \bibinfo {author} {\bibfnamefont {Y.}~\bibnamefont {Zhang}}, \bibinfo {author} {\bibfnamefont {M.~T.}\ \bibnamefont {Edmonds}}, \bibinfo {author} {\bibfnamefont {S.}~\bibnamefont {Li}}, \bibinfo {author} {\bibfnamefont {J.}~\bibnamefont {Seidel}}, \bibinfo {author} {\bibfnamefont {Y.}~\bibnamefont {Zhu}}, \bibinfo {author} {\bibfnamefont {J.~Z.}\ \bibnamefont {Liu}}, \bibinfo {author} {\bibfnamefont {W.-X.}\ \bibnamefont {Tang}},\ and\ \bibinfo {author}
  {\bibfnamefont {M.~S.}\ \bibnamefont {Fuhrer}},\ }\bibfield  {title} {\bibinfo {title} {Room temperature in-plane ferroelectricity in van der waals {In$_2$Se$_3$}},\ }\href {https://doi.org/10.1126/sciadv.aar7720} {\bibfield  {journal} {\bibinfo  {journal} {Sci. Adv.}\ }\textbf {\bibinfo {volume} {4}},\ \bibinfo {pages} {eaar7720} (\bibinfo {year} {2018})}\BibitemShut {NoStop}%
\bibitem [{\citenamefont {Zhang}\ \emph {et~al.}(2021)\citenamefont {Zhang}, \citenamefont {Nie}, \citenamefont {Zhang}, \citenamefont {Yuan}, \citenamefont {Fu},\ and\ \citenamefont {Zhang}}]{Zhang21p2106951}%
  \BibitemOpen
  \bibfield  {author} {\bibinfo {author} {\bibfnamefont {Z.}~\bibnamefont {Zhang}}, \bibinfo {author} {\bibfnamefont {J.}~\bibnamefont {Nie}}, \bibinfo {author} {\bibfnamefont {Z.}~\bibnamefont {Zhang}}, \bibinfo {author} {\bibfnamefont {Y.}~\bibnamefont {Yuan}}, \bibinfo {author} {\bibfnamefont {Y.}~\bibnamefont {Fu}},\ and\ \bibinfo {author} {\bibfnamefont {W.}~\bibnamefont {Zhang}},\ }\bibfield  {title} {\bibinfo {title} {Atomic visualization and switching of ferroelectric order in $\beta$‐{In$_2$Se$_3$} films at the single layer limit},\ }\href {https://doi.org/10.1002/adma.202106951} {\bibfield  {journal} {\bibinfo  {journal} {Adv. Mater.}\ }\textbf {\bibinfo {volume} {34}},\ \bibinfo {pages} {2106951} (\bibinfo {year} {2021})}\BibitemShut {NoStop}%
\bibitem [{\citenamefont {Zhang}\ \emph {et~al.}(2019)\citenamefont {Zhang}, \citenamefont {Wang}, \citenamefont {Dong}, \citenamefont {Nie}, \citenamefont {Xiang}, \citenamefont {Zhu}, \citenamefont {Liu},\ and\ \citenamefont {Tao}}]{Zhang19p8004}%
  \BibitemOpen
  \bibfield  {author} {\bibinfo {author} {\bibfnamefont {F.}~\bibnamefont {Zhang}}, \bibinfo {author} {\bibfnamefont {Z.}~\bibnamefont {Wang}}, \bibinfo {author} {\bibfnamefont {J.}~\bibnamefont {Dong}}, \bibinfo {author} {\bibfnamefont {A.}~\bibnamefont {Nie}}, \bibinfo {author} {\bibfnamefont {J.}~\bibnamefont {Xiang}}, \bibinfo {author} {\bibfnamefont {W.}~\bibnamefont {Zhu}}, \bibinfo {author} {\bibfnamefont {Z.}~\bibnamefont {Liu}},\ and\ \bibinfo {author} {\bibfnamefont {C.}~\bibnamefont {Tao}},\ }\bibfield  {title} {\bibinfo {title} {Atomic-scale observation of reversible thermally driven phase transformation in 2{D} {In$_2$Se$_3$}},\ }\href {https://doi.org/10.1021/acsnano.9b02764} {\bibfield  {journal} {\bibinfo  {journal} {{ACS} Nano}\ }\textbf {\bibinfo {volume} {13}},\ \bibinfo {pages} {8004} (\bibinfo {year} {2019})}\BibitemShut {NoStop}%
\bibitem [{\citenamefont {Zhou}\ \emph {et~al.}(2017)\citenamefont {Zhou}, \citenamefont {Wu}, \citenamefont {Zhu}, \citenamefont {Cho}, \citenamefont {He}, \citenamefont {Yang}, \citenamefont {Herrera}, \citenamefont {Chu}, \citenamefont {Han}, \citenamefont {Downer}, \citenamefont {Peng},\ and\ \citenamefont {Lai}}]{Zhou17p5508}%
  \BibitemOpen
  \bibfield  {author} {\bibinfo {author} {\bibfnamefont {Y.}~\bibnamefont {Zhou}}, \bibinfo {author} {\bibfnamefont {D.}~\bibnamefont {Wu}}, \bibinfo {author} {\bibfnamefont {Y.}~\bibnamefont {Zhu}}, \bibinfo {author} {\bibfnamefont {Y.}~\bibnamefont {Cho}}, \bibinfo {author} {\bibfnamefont {Q.}~\bibnamefont {He}}, \bibinfo {author} {\bibfnamefont {X.}~\bibnamefont {Yang}}, \bibinfo {author} {\bibfnamefont {K.}~\bibnamefont {Herrera}}, \bibinfo {author} {\bibfnamefont {Z.}~\bibnamefont {Chu}}, \bibinfo {author} {\bibfnamefont {Y.}~\bibnamefont {Han}}, \bibinfo {author} {\bibfnamefont {M.~C.}\ \bibnamefont {Downer}}, \bibinfo {author} {\bibfnamefont {H.}~\bibnamefont {Peng}},\ and\ \bibinfo {author} {\bibfnamefont {K.}~\bibnamefont {Lai}},\ }\bibfield  {title} {\bibinfo {title} {Out-of-plane piezoelectricity and ferroelectricity in layered {$\alpha$-In$_2$Se$_3$} nanoflakes},\ }\href {https://doi.org/10.1021/acs.nanolett.7b02198} {\bibfield  {journal} {\bibinfo  {journal} {Nano Lett.}\ }\textbf {\bibinfo
  {volume} {17}},\ \bibinfo {pages} {5508} (\bibinfo {year} {2017})}\BibitemShut {NoStop}%
\bibitem [{\citenamefont {Xue}\ \emph {et~al.}(2018)\citenamefont {Xue}, \citenamefont {Hu}, \citenamefont {Lee}, \citenamefont {Lu}, \citenamefont {Zhang}, \citenamefont {Tang}, \citenamefont {Han}, \citenamefont {Hsu}, \citenamefont {Tu}, \citenamefont {Chang}, \citenamefont {Lien}, \citenamefont {He}, \citenamefont {Zhang}, \citenamefont {Li},\ and\ \citenamefont {Zhang}}]{Xue18p1803738}%
  \BibitemOpen
  \bibfield  {author} {\bibinfo {author} {\bibfnamefont {F.}~\bibnamefont {Xue}}, \bibinfo {author} {\bibfnamefont {W.}~\bibnamefont {Hu}}, \bibinfo {author} {\bibfnamefont {K.-C.}\ \bibnamefont {Lee}}, \bibinfo {author} {\bibfnamefont {L.-S.}\ \bibnamefont {Lu}}, \bibinfo {author} {\bibfnamefont {J.}~\bibnamefont {Zhang}}, \bibinfo {author} {\bibfnamefont {H.-L.}\ \bibnamefont {Tang}}, \bibinfo {author} {\bibfnamefont {A.}~\bibnamefont {Han}}, \bibinfo {author} {\bibfnamefont {W.-T.}\ \bibnamefont {Hsu}}, \bibinfo {author} {\bibfnamefont {S.}~\bibnamefont {Tu}}, \bibinfo {author} {\bibfnamefont {W.-H.}\ \bibnamefont {Chang}}, \bibinfo {author} {\bibfnamefont {C.-H.}\ \bibnamefont {Lien}}, \bibinfo {author} {\bibfnamefont {J.-H.}\ \bibnamefont {He}}, \bibinfo {author} {\bibfnamefont {Z.}~\bibnamefont {Zhang}}, \bibinfo {author} {\bibfnamefont {L.-J.}\ \bibnamefont {Li}},\ and\ \bibinfo {author} {\bibfnamefont {X.}~\bibnamefont {Zhang}},\ }\bibfield  {title} {\bibinfo {title} {Room-temperature ferroelectricity
  in hexagonally layered {$\alpha$-In$_2$Se$_3$} nanoflakes down to the monolayer limit},\ }\href {https://doi.org/10.1002/adfm.201803738} {\bibfield  {journal} {\bibinfo  {journal} {Adv. Funct. Mater.}\ }\textbf {\bibinfo {volume} {28}},\ \bibinfo {pages} {1803738} (\bibinfo {year} {2018})}\BibitemShut {NoStop}%
\bibitem [{\citenamefont {Cui}\ \emph {et~al.}(2018)\citenamefont {Cui}, \citenamefont {Hu}, \citenamefont {Yan}, \citenamefont {Addiego}, \citenamefont {Gao}, \citenamefont {Wang}, \citenamefont {Wang}, \citenamefont {Li}, \citenamefont {Cheng}, \citenamefont {Li}, \citenamefont {Zhang}, \citenamefont {Alshareef}, \citenamefont {Wu}, \citenamefont {Zhu}, \citenamefont {Pan},\ and\ \citenamefont {Li}}]{Cui18p1253}%
  \BibitemOpen
  \bibfield  {author} {\bibinfo {author} {\bibfnamefont {C.}~\bibnamefont {Cui}}, \bibinfo {author} {\bibfnamefont {W.-J.}\ \bibnamefont {Hu}}, \bibinfo {author} {\bibfnamefont {X.}~\bibnamefont {Yan}}, \bibinfo {author} {\bibfnamefont {C.}~\bibnamefont {Addiego}}, \bibinfo {author} {\bibfnamefont {W.}~\bibnamefont {Gao}}, \bibinfo {author} {\bibfnamefont {Y.}~\bibnamefont {Wang}}, \bibinfo {author} {\bibfnamefont {Z.}~\bibnamefont {Wang}}, \bibinfo {author} {\bibfnamefont {L.}~\bibnamefont {Li}}, \bibinfo {author} {\bibfnamefont {Y.}~\bibnamefont {Cheng}}, \bibinfo {author} {\bibfnamefont {P.}~\bibnamefont {Li}}, \bibinfo {author} {\bibfnamefont {X.}~\bibnamefont {Zhang}}, \bibinfo {author} {\bibfnamefont {H.~N.}\ \bibnamefont {Alshareef}}, \bibinfo {author} {\bibfnamefont {T.}~\bibnamefont {Wu}}, \bibinfo {author} {\bibfnamefont {W.}~\bibnamefont {Zhu}}, \bibinfo {author} {\bibfnamefont {X.}~\bibnamefont {Pan}},\ and\ \bibinfo {author} {\bibfnamefont {L.-J.}\ \bibnamefont {Li}},\ }\bibfield  {title}
  {\bibinfo {title} {Intercorrelated in-plane and out-of-plane ferroelectricity in ultrathin two-dimensional layered semiconductor {In$_2$Se$_3$}},\ }\href {https://doi.org/10.1021/acs.nanolett.7b04852} {\bibfield  {journal} {\bibinfo  {journal} {Nano Lett.}\ }\textbf {\bibinfo {volume} {18}},\ \bibinfo {pages} {1253} (\bibinfo {year} {2018})}\BibitemShut {NoStop}%
\bibitem [{\citenamefont {Wan}\ \emph {et~al.}(2018)\citenamefont {Wan}, \citenamefont {Li}, \citenamefont {Li}, \citenamefont {Mao}, \citenamefont {Zhu},\ and\ \citenamefont {Zeng}}]{Wan18p14885}%
  \BibitemOpen
  \bibfield  {author} {\bibinfo {author} {\bibfnamefont {S.}~\bibnamefont {Wan}}, \bibinfo {author} {\bibfnamefont {Y.}~\bibnamefont {Li}}, \bibinfo {author} {\bibfnamefont {W.}~\bibnamefont {Li}}, \bibinfo {author} {\bibfnamefont {X.}~\bibnamefont {Mao}}, \bibinfo {author} {\bibfnamefont {W.}~\bibnamefont {Zhu}},\ and\ \bibinfo {author} {\bibfnamefont {H.}~\bibnamefont {Zeng}},\ }\bibfield  {title} {\bibinfo {title} {Room-temperature ferroelectricity and a switchable diode effect in two-dimensional {$\alpha$-In$_2$Se$_3$} thin layers},\ }\href {https://doi.org/10.1039/c8nr04422h} {\bibfield  {journal} {\bibinfo  {journal} {Nanoscale}\ }\textbf {\bibinfo {volume} {10}},\ \bibinfo {pages} {14885} (\bibinfo {year} {2018})}\BibitemShut {NoStop}%
\bibitem [{\citenamefont {Togo}\ \emph {et~al.}(2015)\citenamefont {Togo}, \citenamefont {Chaput},\ and\ \citenamefont {Tanaka}}]{Togo15p094306}%
  \BibitemOpen
  \bibfield  {author} {\bibinfo {author} {\bibfnamefont {A.}~\bibnamefont {Togo}}, \bibinfo {author} {\bibfnamefont {L.}~\bibnamefont {Chaput}},\ and\ \bibinfo {author} {\bibfnamefont {I.}~\bibnamefont {Tanaka}},\ }\bibfield  {title} {\bibinfo {title} {Distributions of phonon lifetimes in brillouin zones},\ }\href {https://doi.org/10.1103/PhysRevB.91.094306} {\bibfield  {journal} {\bibinfo  {journal} {Phys. Rev. B}\ }\textbf {\bibinfo {volume} {91}},\ \bibinfo {pages} {094306} (\bibinfo {year} {2015})}\BibitemShut {NoStop}%
\bibitem [{\citenamefont {Togo}\ \emph {et~al.}(2023)\citenamefont {Togo}, \citenamefont {Chaput}, \citenamefont {Tadano},\ and\ \citenamefont {Tanaka}}]{Togo23p353001}%
  \BibitemOpen
  \bibfield  {author} {\bibinfo {author} {\bibfnamefont {A.}~\bibnamefont {Togo}}, \bibinfo {author} {\bibfnamefont {L.}~\bibnamefont {Chaput}}, \bibinfo {author} {\bibfnamefont {T.}~\bibnamefont {Tadano}},\ and\ \bibinfo {author} {\bibfnamefont {I.}~\bibnamefont {Tanaka}},\ }\bibfield  {title} {\bibinfo {title} {Implementation strategies in phonopy and phono3py},\ }\href {https://doi.org/10.1088/1361-648X/acd831} {\bibfield  {journal} {\bibinfo  {journal} {J. Phys. Condens. Matter}\ }\textbf {\bibinfo {volume} {35}},\ \bibinfo {pages} {353001} (\bibinfo {year} {2023})}\BibitemShut {NoStop}%
\bibitem [{\citenamefont {Chaput}(2013)}]{Chaput13p265506}%
  \BibitemOpen
  \bibfield  {author} {\bibinfo {author} {\bibfnamefont {L.}~\bibnamefont {Chaput}},\ }\bibfield  {title} {\bibinfo {title} {Direct solution to the linearized phonon boltzmann equation},\ }\href {https://doi.org/10.1103/PhysRevLett.110.265506} {\bibfield  {journal} {\bibinfo  {journal} {Phys. Rev. Lett.}\ }\textbf {\bibinfo {volume} {110}},\ \bibinfo {pages} {265506} (\bibinfo {year} {2013})}\BibitemShut {NoStop}%
\bibitem [{\citenamefont {Simoncelli}\ \emph {et~al.}(2022)\citenamefont {Simoncelli}, \citenamefont {Marzari},\ and\ \citenamefont {Mauri}}]{Simoncelli22p041011}%
  \BibitemOpen
  \bibfield  {author} {\bibinfo {author} {\bibfnamefont {M.}~\bibnamefont {Simoncelli}}, \bibinfo {author} {\bibfnamefont {N.}~\bibnamefont {Marzari}},\ and\ \bibinfo {author} {\bibfnamefont {F.}~\bibnamefont {Mauri}},\ }\bibfield  {title} {\bibinfo {title} {Wigner formulation of thermal transport in solids},\ }\href {https://doi.org/10.1103/PhysRevX.12.041011} {\bibfield  {journal} {\bibinfo  {journal} {Phys. Rev. X}\ }\textbf {\bibinfo {volume} {12}},\ \bibinfo {pages} {041011} (\bibinfo {year} {2022})}\BibitemShut {NoStop}%
\bibitem [{\citenamefont {Fan}\ \emph {et~al.}(2021)\citenamefont {Fan}, \citenamefont {Zeng}, \citenamefont {Zhang}, \citenamefont {Wang}, \citenamefont {Song}, \citenamefont {Dong}, \citenamefont {Chen},\ and\ \citenamefont {Ala-Nissila}}]{Fan21p104309}%
  \BibitemOpen
  \bibfield  {author} {\bibinfo {author} {\bibfnamefont {Z.}~\bibnamefont {Fan}}, \bibinfo {author} {\bibfnamefont {Z.}~\bibnamefont {Zeng}}, \bibinfo {author} {\bibfnamefont {C.}~\bibnamefont {Zhang}}, \bibinfo {author} {\bibfnamefont {Y.}~\bibnamefont {Wang}}, \bibinfo {author} {\bibfnamefont {K.}~\bibnamefont {Song}}, \bibinfo {author} {\bibfnamefont {H.}~\bibnamefont {Dong}}, \bibinfo {author} {\bibfnamefont {Y.}~\bibnamefont {Chen}},\ and\ \bibinfo {author} {\bibfnamefont {T.}~\bibnamefont {Ala-Nissila}},\ }\bibfield  {title} {\bibinfo {title} {Neuroevolution machine learning potentials: Combining high accuracy and low cost in atomistic simulations and application to heat transport},\ }\href {https://doi.org/10.1103/PhysRevB.104.104309} {\bibfield  {journal} {\bibinfo  {journal} {Phys. Rev. B}\ }\textbf {\bibinfo {volume} {104}},\ \bibinfo {pages} {104309} (\bibinfo {year} {2021})}\BibitemShut {NoStop}%
\bibitem [{\citenamefont {Fan}\ \emph {et~al.}(2022)\citenamefont {Fan}, \citenamefont {Wang}, \citenamefont {Ying}, \citenamefont {Song}, \citenamefont {Wang}, \citenamefont {Wang}, \citenamefont {Zeng}, \citenamefont {Xu}, \citenamefont {Lindgren}, \citenamefont {Rahm}, \citenamefont {Gabourie}, \citenamefont {Liu}, \citenamefont {Dong}, \citenamefont {Wu}, \citenamefont {Chen}, \citenamefont {Zhong}, \citenamefont {Sun}, \citenamefont {Erhart}, \citenamefont {Su},\ and\ \citenamefont {Ala-Nissila}}]{Fan22p114801}%
  \BibitemOpen
  \bibfield  {author} {\bibinfo {author} {\bibfnamefont {Z.}~\bibnamefont {Fan}}, \bibinfo {author} {\bibfnamefont {Y.}~\bibnamefont {Wang}}, \bibinfo {author} {\bibfnamefont {P.}~\bibnamefont {Ying}}, \bibinfo {author} {\bibfnamefont {K.}~\bibnamefont {Song}}, \bibinfo {author} {\bibfnamefont {J.}~\bibnamefont {Wang}}, \bibinfo {author} {\bibfnamefont {Y.}~\bibnamefont {Wang}}, \bibinfo {author} {\bibfnamefont {Z.}~\bibnamefont {Zeng}}, \bibinfo {author} {\bibfnamefont {K.}~\bibnamefont {Xu}}, \bibinfo {author} {\bibfnamefont {E.}~\bibnamefont {Lindgren}}, \bibinfo {author} {\bibfnamefont {J.~M.}\ \bibnamefont {Rahm}}, \bibinfo {author} {\bibfnamefont {A.~J.}\ \bibnamefont {Gabourie}}, \bibinfo {author} {\bibfnamefont {J.}~\bibnamefont {Liu}}, \bibinfo {author} {\bibfnamefont {H.}~\bibnamefont {Dong}}, \bibinfo {author} {\bibfnamefont {J.}~\bibnamefont {Wu}}, \bibinfo {author} {\bibfnamefont {Y.}~\bibnamefont {Chen}}, \bibinfo {author} {\bibfnamefont {Z.}~\bibnamefont {Zhong}}, \bibinfo {author}
  {\bibfnamefont {J.}~\bibnamefont {Sun}}, \bibinfo {author} {\bibfnamefont {P.}~\bibnamefont {Erhart}}, \bibinfo {author} {\bibfnamefont {Y.}~\bibnamefont {Su}},\ and\ \bibinfo {author} {\bibfnamefont {T.}~\bibnamefont {Ala-Nissila}},\ }\bibfield  {title} {\bibinfo {title} {{GPUMD}: A package for constructing accurate machine-learned potentials and performing highly efficient atomistic simulations},\ }\href {https://doi.org/10.1063/5.0106617} {\bibfield  {journal} {\bibinfo  {journal} {J. Chem. Phys.}\ }\textbf {\bibinfo {volume} {157}},\ \bibinfo {pages} {114801} (\bibinfo {year} {2022})}\BibitemShut {NoStop}%
\bibitem [{\citenamefont {Sonti}\ \emph {et~al.}(2024)\citenamefont {Sonti}, \citenamefont {Sun}, \citenamefont {Chen}, \citenamefont {Kowalski}, \citenamefont {Kowalski}, \citenamefont {Donadio}, \citenamefont {Ahn},\ and\ \citenamefont {Kulkarni}}]{Sonti24p8261}%
  \BibitemOpen
  \bibfield  {author} {\bibinfo {author} {\bibfnamefont {S.}~\bibnamefont {Sonti}}, \bibinfo {author} {\bibfnamefont {C.}~\bibnamefont {Sun}}, \bibinfo {author} {\bibfnamefont {Z.}~\bibnamefont {Chen}}, \bibinfo {author} {\bibfnamefont {R.~M.}\ \bibnamefont {Kowalski}}, \bibinfo {author} {\bibfnamefont {J.~S.}\ \bibnamefont {Kowalski}}, \bibinfo {author} {\bibfnamefont {D.}~\bibnamefont {Donadio}}, \bibinfo {author} {\bibfnamefont {S.-H.}\ \bibnamefont {Ahn}},\ and\ \bibinfo {author} {\bibfnamefont {A.~R.}\ \bibnamefont {Kulkarni}},\ }\bibfield  {title} {\bibinfo {title} {Stability and dynamics of zeolite-confined gold nanoclusters},\ }\href {https://doi.org/10.1021/acs.jctc.4c00978} {\bibfield  {journal} {\bibinfo  {journal} {J. Chem. Theory Comput.}\ }\textbf {\bibinfo {volume} {20}},\ \bibinfo {pages} {8261} (\bibinfo {year} {2024})}\BibitemShut {NoStop}%
\bibitem [{\citenamefont {Bl{\"o}chl}(1994)}]{Blochl94p17953}%
  \BibitemOpen
  \bibfield  {author} {\bibinfo {author} {\bibfnamefont {P.~E.}\ \bibnamefont {Bl{\"o}chl}},\ }\bibfield  {title} {\bibinfo {title} {Projector augmented-wave method},\ }\href {https://doi.org/10.1103/PhysRevB.50.17953} {\bibfield  {journal} {\bibinfo  {journal} {Phys. Rev. B}\ }\textbf {\bibinfo {volume} {50}},\ \bibinfo {pages} {17953} (\bibinfo {year} {1994})}\BibitemShut {NoStop}%
\bibitem [{\citenamefont {Kresse}\ and\ \citenamefont {Furthm{\"u}ller}(1996)}]{Kresse96p11169}%
  \BibitemOpen
  \bibfield  {author} {\bibinfo {author} {\bibfnamefont {G.}~\bibnamefont {Kresse}}\ and\ \bibinfo {author} {\bibfnamefont {J.}~\bibnamefont {Furthm{\"u}ller}},\ }\bibfield  {title} {\bibinfo {title} {Efficient iterative schemes for \textit{ab initio} total-energy calculations using a plane-wave basis set},\ }\href {https://doi.org/10.1103/PhysRevB.54.11169} {\bibfield  {journal} {\bibinfo  {journal} {Phys. Rev. B}\ }\textbf {\bibinfo {volume} {54}},\ \bibinfo {pages} {11169} (\bibinfo {year} {1996})}\BibitemShut {NoStop}%
\bibitem [{\citenamefont {Kresse}\ and\ \citenamefont {Furthm\"{u}ller}(1996)}]{Kresse96p15}%
  \BibitemOpen
  \bibfield  {author} {\bibinfo {author} {\bibfnamefont {G.}~\bibnamefont {Kresse}}\ and\ \bibinfo {author} {\bibfnamefont {J.}~\bibnamefont {Furthm\"{u}ller}},\ }\bibfield  {title} {\bibinfo {title} {Efficiency of ab-initio total energy calculations for metals and semiconductors using a plane-wave basis set},\ }\href {https://doi.org/10.1016/0927-0256(96)00008-0} {\bibfield  {journal} {\bibinfo  {journal} {Comput. Mater. Sci.}\ }\textbf {\bibinfo {volume} {6}},\ \bibinfo {pages} {15} (\bibinfo {year} {1996})}\BibitemShut {NoStop}%
\bibitem [{\citenamefont {Perdew}\ \emph {et~al.}(2008)\citenamefont {Perdew}, \citenamefont {Ruzsinszky}, \citenamefont {Csonka}, \citenamefont {Vydrov}, \citenamefont {Scuseria}, \citenamefont {Constantin}, \citenamefont {Zhou},\ and\ \citenamefont {Burke}}]{Perdew08p136406}%
  \BibitemOpen
  \bibfield  {author} {\bibinfo {author} {\bibfnamefont {J.~P.}\ \bibnamefont {Perdew}}, \bibinfo {author} {\bibfnamefont {A.}~\bibnamefont {Ruzsinszky}}, \bibinfo {author} {\bibfnamefont {G.~I.}\ \bibnamefont {Csonka}}, \bibinfo {author} {\bibfnamefont {O.~A.}\ \bibnamefont {Vydrov}}, \bibinfo {author} {\bibfnamefont {G.~E.}\ \bibnamefont {Scuseria}}, \bibinfo {author} {\bibfnamefont {L.~A.}\ \bibnamefont {Constantin}}, \bibinfo {author} {\bibfnamefont {X.}~\bibnamefont {Zhou}},\ and\ \bibinfo {author} {\bibfnamefont {K.}~\bibnamefont {Burke}},\ }\bibfield  {title} {\bibinfo {title} {Restoring the density-gradient expansion for exchange in solids and surfaces},\ }\href {https://doi.org/10.1103/PhysRevLett.100.136406} {\bibfield  {journal} {\bibinfo  {journal} {Phys. Rev. Lett.}\ }\textbf {\bibinfo {volume} {100}},\ \bibinfo {pages} {136406} (\bibinfo {year} {2008})}\BibitemShut {NoStop}%
\bibitem [{\citenamefont {Fan}\ \emph {et~al.}(2019)\citenamefont {Fan}, \citenamefont {Dong}, \citenamefont {Harju},\ and\ \citenamefont {Ala-Nissila}}]{Fan19p064308}%
  \BibitemOpen
  \bibfield  {author} {\bibinfo {author} {\bibfnamefont {Z.}~\bibnamefont {Fan}}, \bibinfo {author} {\bibfnamefont {H.}~\bibnamefont {Dong}}, \bibinfo {author} {\bibfnamefont {A.}~\bibnamefont {Harju}},\ and\ \bibinfo {author} {\bibfnamefont {T.}~\bibnamefont {Ala-Nissila}},\ }\bibfield  {title} {\bibinfo {title} {Homogeneous nonequilibrium molecular dynamics method for heat transport and spectral decomposition with many-body potentials},\ }\href {https://doi.org/10.1103/PhysRevB.99.064308} {\bibfield  {journal} {\bibinfo  {journal} {Phys. Rev. B}\ }\textbf {\bibinfo {volume} {99}},\ \bibinfo {pages} {064308} (\bibinfo {year} {2019})}\BibitemShut {NoStop}%
\bibitem [{\citenamefont {Chen}\ \emph {et~al.}(2024)\citenamefont {Chen}, \citenamefont {Berrens}, \citenamefont {Chan}, \citenamefont {Fan},\ and\ \citenamefont {Donadio}}]{Chen24p128}%
  \BibitemOpen
  \bibfield  {author} {\bibinfo {author} {\bibfnamefont {Z.}~\bibnamefont {Chen}}, \bibinfo {author} {\bibfnamefont {M.~L.}\ \bibnamefont {Berrens}}, \bibinfo {author} {\bibfnamefont {K.-T.}\ \bibnamefont {Chan}}, \bibinfo {author} {\bibfnamefont {Z.}~\bibnamefont {Fan}},\ and\ \bibinfo {author} {\bibfnamefont {D.}~\bibnamefont {Donadio}},\ }\bibfield  {title} {\bibinfo {title} {Thermodynamics of water and ice from a fast and scalable first-principles neuroevolution potential},\ }\href {https://doi.org/10.1021/acs.jced.3c00561} {\bibfield  {journal} {\bibinfo  {journal} {J. Chem. Eng. Data}\ }\textbf {\bibinfo {volume} {69}},\ \bibinfo {pages} {128} (\bibinfo {year} {2024})}\BibitemShut {NoStop}%
\bibitem [{\citenamefont {Nian}\ \emph {et~al.}(2021)\citenamefont {Nian}, \citenamefont {Wang},\ and\ \citenamefont {Dong}}]{Nian21p033103}%
  \BibitemOpen
  \bibfield  {author} {\bibinfo {author} {\bibfnamefont {T.}~\bibnamefont {Nian}}, \bibinfo {author} {\bibfnamefont {Z.}~\bibnamefont {Wang}},\ and\ \bibinfo {author} {\bibfnamefont {B.}~\bibnamefont {Dong}},\ }\bibfield  {title} {\bibinfo {title} {Thermoelectric properties of $\alpha$-{In$_2$Se$_3$} monolayer},\ }\href {https://doi.org/10.1063/5.0036316} {\bibfield  {journal} {\bibinfo  {journal} {Appl. Phys. Lett.}\ }\textbf {\bibinfo {volume} {118}},\ \bibinfo {pages} {033103} (\bibinfo {year} {2021})}\BibitemShut {NoStop}%
\bibitem [{\citenamefont {Han}\ \emph {et~al.}(2014)\citenamefont {Han}, \citenamefont {Chen}, \citenamefont {Drennan},\ and\ \citenamefont {Zou}}]{Han14p2747}%
  \BibitemOpen
  \bibfield  {author} {\bibinfo {author} {\bibfnamefont {G.}~\bibnamefont {Han}}, \bibinfo {author} {\bibfnamefont {Z.-G.}\ \bibnamefont {Chen}}, \bibinfo {author} {\bibfnamefont {J.}~\bibnamefont {Drennan}},\ and\ \bibinfo {author} {\bibfnamefont {J.}~\bibnamefont {Zou}},\ }\bibfield  {title} {\bibinfo {title} {Indium selenides: Structural characteristics, synthesis and their thermoelectric performances},\ }\href {https://doi.org/10.1002/smll.201400104} {\bibfield  {journal} {\bibinfo  {journal} {Small}\ }\textbf {\bibinfo {volume} {10}},\ \bibinfo {pages} {2747} (\bibinfo {year} {2014})}\BibitemShut {NoStop}%
\bibitem [{\citenamefont {Hung}\ \emph {et~al.}(2017)\citenamefont {Hung}, \citenamefont {Nugraha},\ and\ \citenamefont {Saito}}]{Hung17p092107}%
  \BibitemOpen
  \bibfield  {author} {\bibinfo {author} {\bibfnamefont {N.~T.}\ \bibnamefont {Hung}}, \bibinfo {author} {\bibfnamefont {A.~R.~T.}\ \bibnamefont {Nugraha}},\ and\ \bibinfo {author} {\bibfnamefont {R.}~\bibnamefont {Saito}},\ }\bibfield  {title} {\bibinfo {title} {Two-dimensional {InSe} as a potential thermoelectric material},\ }\href {https://doi.org/10.1063/1.5001184} {\bibfield  {journal} {\bibinfo  {journal} {Appl. Phys. Lett.}\ }\textbf {\bibinfo {volume} {111}},\ \bibinfo {pages} {092107} (\bibinfo {year} {2017})}\BibitemShut {NoStop}%
\bibitem [{\citenamefont {Qi}\ \emph {et~al.}(2023)\citenamefont {Qi}, \citenamefont {Wu}, \citenamefont {Lu},\ and\ \citenamefont {Liu}}]{Qi23p085701}%
  \BibitemOpen
  \bibfield  {author} {\bibinfo {author} {\bibfnamefont {H.}~\bibnamefont {Qi}}, \bibinfo {author} {\bibfnamefont {C.}~\bibnamefont {Wu}}, \bibinfo {author} {\bibfnamefont {P.}~\bibnamefont {Lu}},\ and\ \bibinfo {author} {\bibfnamefont {C.}~\bibnamefont {Liu}},\ }\bibfield  {title} {\bibinfo {title} {Phonon thermal transport in ferroelectric $\alpha$-{In$_2$Se$_3$} via first-principles calculations},\ }\href {https://doi.org/10.1088/1361-6528/ad0c75} {\bibfield  {journal} {\bibinfo  {journal} {Nanotechnology}\ }\textbf {\bibinfo {volume} {35}},\ \bibinfo {pages} {085701} (\bibinfo {year} {2023})}\BibitemShut {NoStop}%
\bibitem [{\citenamefont {Wu}\ \emph {et~al.}(2021)\citenamefont {Wu}, \citenamefont {Bai}, \citenamefont {Huang}, \citenamefont {Ma}, \citenamefont {Liu},\ and\ \citenamefont {Liu}}]{Wu21p174107}%
  \BibitemOpen
  \bibfield  {author} {\bibinfo {author} {\bibfnamefont {J.}~\bibnamefont {Wu}}, \bibinfo {author} {\bibfnamefont {L.}~\bibnamefont {Bai}}, \bibinfo {author} {\bibfnamefont {J.}~\bibnamefont {Huang}}, \bibinfo {author} {\bibfnamefont {L.}~\bibnamefont {Ma}}, \bibinfo {author} {\bibfnamefont {J.}~\bibnamefont {Liu}},\ and\ \bibinfo {author} {\bibfnamefont {S.}~\bibnamefont {Liu}},\ }\bibfield  {title} {\bibinfo {title} {Accurate force field of two-dimensional ferroelectrics from deep learning},\ }\href {https://doi.org/10.1103/physrevb.104.174107} {\bibfield  {journal} {\bibinfo  {journal} {Phys. Rev. B}\ }\textbf {\bibinfo {volume} {104}},\ \bibinfo {pages} {174107} (\bibinfo {year} {2021})}\BibitemShut {NoStop}%
\bibitem [{\citenamefont {Barbalinardo}\ \emph {et~al.}(2020)\citenamefont {Barbalinardo}, \citenamefont {Chen}, \citenamefont {Lundgren},\ and\ \citenamefont {Donadio}}]{Barbalinardo20p135104}%
  \BibitemOpen
  \bibfield  {author} {\bibinfo {author} {\bibfnamefont {G.}~\bibnamefont {Barbalinardo}}, \bibinfo {author} {\bibfnamefont {Z.}~\bibnamefont {Chen}}, \bibinfo {author} {\bibfnamefont {N.~W.}\ \bibnamefont {Lundgren}},\ and\ \bibinfo {author} {\bibfnamefont {D.}~\bibnamefont {Donadio}},\ }\bibfield  {title} {\bibinfo {title} {Efficient anharmonic lattice dynamics calculations of thermal transport in crystalline and disordered solids},\ }\href {https://doi.org/10.1063/5.0020443} {\bibfield  {journal} {\bibinfo  {journal} {J. Appl. Phys.}\ }\textbf {\bibinfo {volume} {128}},\ \bibinfo {pages} {135104} (\bibinfo {year} {2020})}\BibitemShut {NoStop}%
\bibitem [{\citenamefont {Tao}\ and\ \citenamefont {Gu}(2013)}]{Tao13p3501}%
  \BibitemOpen
  \bibfield  {author} {\bibinfo {author} {\bibfnamefont {X.}~\bibnamefont {Tao}}\ and\ \bibinfo {author} {\bibfnamefont {Y.}~\bibnamefont {Gu}},\ }\bibfield  {title} {\bibinfo {title} {{Crystalline–Crystalline} phase transformation in two-dimensional {In$_2$Se$_3$} thin layers},\ }\href {https://doi.org/10.1021/nl400888p} {\bibfield  {journal} {\bibinfo  {journal} {Nano Lett.}\ }\textbf {\bibinfo {volume} {13}},\ \bibinfo {pages} {3501–3505} (\bibinfo {year} {2013})}\BibitemShut {NoStop}%
\bibitem [{\citenamefont {Camiola}\ \emph {et~al.}(2023)\citenamefont {Camiola}, \citenamefont {Romano},\ and\ \citenamefont {Vitanza}}]{Camiola23p10}%
  \BibitemOpen
  \bibfield  {author} {\bibinfo {author} {\bibfnamefont {V.~D.}\ \bibnamefont {Camiola}}, \bibinfo {author} {\bibfnamefont {V.}~\bibnamefont {Romano}},\ and\ \bibinfo {author} {\bibfnamefont {G.}~\bibnamefont {Vitanza}},\ }\bibfield  {title} {\bibinfo {title} {Wigner equations for phonons transport and quantum heat flux},\ }\href {https://doi.org/10.1007/s00332-023-09993-z} {\bibfield  {journal} {\bibinfo  {journal} {J. Nonlinear Sci.}\ }\textbf {\bibinfo {volume} {34}},\ \bibinfo {pages} {10} (\bibinfo {year} {2023})}\BibitemShut {NoStop}%
\bibitem [{\citenamefont {Das}\ and\ \citenamefont {Banerji}(2023)}]{Das23p11521}%
  \BibitemOpen
  \bibfield  {author} {\bibinfo {author} {\bibfnamefont {A.}~\bibnamefont {Das}}\ and\ \bibinfo {author} {\bibfnamefont {P.}~\bibnamefont {Banerji}},\ }\bibfield  {title} {\bibinfo {title} {Antibonding or nonbonding interaction-driven phonon modes softening and wave-like interband thermal conduction in layered {In$_4$Te$_3$} under the framework of wigner transport formalism},\ }\href {https://doi.org/10.1021/acsaem.3c01838} {\bibfield  {journal} {\bibinfo  {journal} {ACS Appl. Energy Mater.}\ }\textbf {\bibinfo {volume} {6}},\ \bibinfo {pages} {11521} (\bibinfo {year} {2023})}\BibitemShut {NoStop}%
\bibitem [{\citenamefont {Di~Lucente}\ \emph {et~al.}(2023)\citenamefont {Di~Lucente}, \citenamefont {Simoncelli},\ and\ \citenamefont {Marzari}}]{Lucente23p033125}%
  \BibitemOpen
  \bibfield  {author} {\bibinfo {author} {\bibfnamefont {E.}~\bibnamefont {Di~Lucente}}, \bibinfo {author} {\bibfnamefont {M.}~\bibnamefont {Simoncelli}},\ and\ \bibinfo {author} {\bibfnamefont {N.}~\bibnamefont {Marzari}},\ }\bibfield  {title} {\bibinfo {title} {Crossover from boltzmann to wigner thermal transport in thermoelectric skutterudites},\ }\href {https://doi.org/10.1103/PhysRevResearch.5.033125} {\bibfield  {journal} {\bibinfo  {journal} {Phys. Rev. Res.}\ }\textbf {\bibinfo {volume} {5}},\ \bibinfo {pages} {033125} (\bibinfo {year} {2023})}\BibitemShut {NoStop}%
\bibitem [{\citenamefont {Pazhedath}\ \emph {et~al.}(2024)\citenamefont {Pazhedath}, \citenamefont {Bastonero}, \citenamefont {Marzari},\ and\ \citenamefont {Simoncelli}}]{Pazhedath24p024064}%
  \BibitemOpen
  \bibfield  {author} {\bibinfo {author} {\bibfnamefont {A.}~\bibnamefont {Pazhedath}}, \bibinfo {author} {\bibfnamefont {L.}~\bibnamefont {Bastonero}}, \bibinfo {author} {\bibfnamefont {N.}~\bibnamefont {Marzari}},\ and\ \bibinfo {author} {\bibfnamefont {M.}~\bibnamefont {Simoncelli}},\ }\bibfield  {title} {\bibinfo {title} {First-principles characterization of thermal conductivity in {LaPO$_4$}-based alloys},\ }\href {https://doi.org/10.1103/physrevapplied.22.024064} {\bibfield  {journal} {\bibinfo  {journal} {Phys. Rev. Appl.}\ }\textbf {\bibinfo {volume} {22}},\ \bibinfo {pages} {024064} (\bibinfo {year} {2024})}\BibitemShut {NoStop}%
\bibitem [{\citenamefont {Liu}\ and\ \citenamefont {Pantelides}(2019)}]{Liu19p025001}%
  \BibitemOpen
  \bibfield  {author} {\bibinfo {author} {\bibfnamefont {J.}~\bibnamefont {Liu}}\ and\ \bibinfo {author} {\bibfnamefont {S.~T.}\ \bibnamefont {Pantelides}},\ }\bibfield  {title} {\bibinfo {title} {Pyroelectric response and temperature-induced $\alpha$-$\beta$ phase transitions in $\alpha$-{In$_2$Se$_3$} and other $\alpha$-{III$_2$VI$_3$} ({III}= {Al}, {Ga}, {In}; {VI}= {S}, {Se}) monolayers},\ }\href {https://doi.org/10.1088/2053-1583/aaf946} {\bibfield  {journal} {\bibinfo  {journal} {2D Mater.}\ }\textbf {\bibinfo {volume} {6}},\ \bibinfo {pages} {025001} (\bibinfo {year} {2019})}\BibitemShut {NoStop}%
\bibitem [{\citenamefont {Xu}\ \emph {et~al.}(2020)\citenamefont {Xu}, \citenamefont {Chen}, \citenamefont {Cai}, \citenamefont {Meingast}, \citenamefont {Guo}, \citenamefont {Wang}, \citenamefont {Lin}, \citenamefont {Lo}, \citenamefont {Maunders}, \citenamefont {Lazar}, \citenamefont {Wang}, \citenamefont {Lei}, \citenamefont {Chai}, \citenamefont {Zhai}, \citenamefont {Luo},\ and\ \citenamefont {Zhu}}]{Xu20p047601}%
  \BibitemOpen
  \bibfield  {author} {\bibinfo {author} {\bibfnamefont {C.}~\bibnamefont {Xu}}, \bibinfo {author} {\bibfnamefont {Y.}~\bibnamefont {Chen}}, \bibinfo {author} {\bibfnamefont {X.}~\bibnamefont {Cai}}, \bibinfo {author} {\bibfnamefont {A.}~\bibnamefont {Meingast}}, \bibinfo {author} {\bibfnamefont {X.}~\bibnamefont {Guo}}, \bibinfo {author} {\bibfnamefont {F.}~\bibnamefont {Wang}}, \bibinfo {author} {\bibfnamefont {Z.}~\bibnamefont {Lin}}, \bibinfo {author} {\bibfnamefont {T.~W.}\ \bibnamefont {Lo}}, \bibinfo {author} {\bibfnamefont {C.}~\bibnamefont {Maunders}}, \bibinfo {author} {\bibfnamefont {S.}~\bibnamefont {Lazar}}, \bibinfo {author} {\bibfnamefont {N.}~\bibnamefont {Wang}}, \bibinfo {author} {\bibfnamefont {D.}~\bibnamefont {Lei}}, \bibinfo {author} {\bibfnamefont {Y.}~\bibnamefont {Chai}}, \bibinfo {author} {\bibfnamefont {T.}~\bibnamefont {Zhai}}, \bibinfo {author} {\bibfnamefont {X.}~\bibnamefont {Luo}},\ and\ \bibinfo {author} {\bibfnamefont {Y.}~\bibnamefont {Zhu}},\ }\bibfield  {title} {\bibinfo
  {title} {Two-dimensional antiferroelectricity in nanostripe-ordered {In$_2$Se$_3$}},\ }\href {https://doi.org/10.1103/physrevlett.125.047601} {\bibfield  {journal} {\bibinfo  {journal} {Phys. Rev. Lett.}\ }\textbf {\bibinfo {volume} {125}},\ \bibinfo {pages} {047601} (\bibinfo {year} {2020})}\BibitemShut {NoStop}%
\bibitem [{\citenamefont {Umari}\ and\ \citenamefont {Pasquarello}(2002)}]{Umari02p157602}%
  \BibitemOpen
  \bibfield  {author} {\bibinfo {author} {\bibfnamefont {P.}~\bibnamefont {Umari}}\ and\ \bibinfo {author} {\bibfnamefont {A.}~\bibnamefont {Pasquarello}},\ }\bibfield  {title} {\bibinfo {title} {\textit{Ab initio} molecular dynamics in a finite homogeneous electric field},\ }\href {https://doi.org/10.1103/PhysRevLett.89.157602} {\bibfield  {journal} {\bibinfo  {journal} {Phys. Rev. Lett.}\ }\textbf {\bibinfo {volume} {89}},\ \bibinfo {pages} {157602} (\bibinfo {year} {2002})}\BibitemShut {NoStop}%
\bibitem [{\citenamefont {Liu}\ \emph {et~al.}(2016)\citenamefont {Liu}, \citenamefont {Grinberg},\ and\ \citenamefont {Rappe}}]{Liu16p360}%
  \BibitemOpen
  \bibfield  {author} {\bibinfo {author} {\bibfnamefont {S.}~\bibnamefont {Liu}}, \bibinfo {author} {\bibfnamefont {I.}~\bibnamefont {Grinberg}},\ and\ \bibinfo {author} {\bibfnamefont {A.~M.}\ \bibnamefont {Rappe}},\ }\bibfield  {title} {\bibinfo {title} {Intrinsic ferroelectric switching from first principles},\ }\href {https://doi.org/10.1038/nature18286} {\bibfield  {journal} {\bibinfo  {journal} {Nature}\ }\textbf {\bibinfo {volume} {534}},\ \bibinfo {pages} {360} (\bibinfo {year} {2016})}\BibitemShut {NoStop}%
\bibitem [{\citenamefont {Carreras}\ \emph {et~al.}(2017)\citenamefont {Carreras}, \citenamefont {Togo},\ and\ \citenamefont {Tanaka}}]{Carreras17p221}%
  \BibitemOpen
  \bibfield  {author} {\bibinfo {author} {\bibfnamefont {A.}~\bibnamefont {Carreras}}, \bibinfo {author} {\bibfnamefont {A.}~\bibnamefont {Togo}},\ and\ \bibinfo {author} {\bibfnamefont {I.}~\bibnamefont {Tanaka}},\ }\bibfield  {title} {\bibinfo {title} {Dynaphopy: A code for extracting phonon quasiparticles from molecular dynamics simulations},\ }\href {https://doi.org/https://doi.org/10.1016/j.cpc.2017.08.017} {\bibfield  {journal} {\bibinfo  {journal} {Comput. Phys. Commun.}\ }\textbf {\bibinfo {volume} {221}},\ \bibinfo {pages} {221} (\bibinfo {year} {2017})}\BibitemShut {NoStop}%
\bibitem [{\citenamefont {Zeng}\ \emph {et~al.}(2021)\citenamefont {Zeng}, \citenamefont {Zhang}, \citenamefont {Xia}, \citenamefont {Fan}, \citenamefont {Wolverton},\ and\ \citenamefont {Chen}}]{Zeng21p224307}%
  \BibitemOpen
  \bibfield  {author} {\bibinfo {author} {\bibfnamefont {Z.}~\bibnamefont {Zeng}}, \bibinfo {author} {\bibfnamefont {C.}~\bibnamefont {Zhang}}, \bibinfo {author} {\bibfnamefont {Y.}~\bibnamefont {Xia}}, \bibinfo {author} {\bibfnamefont {Z.}~\bibnamefont {Fan}}, \bibinfo {author} {\bibfnamefont {C.}~\bibnamefont {Wolverton}},\ and\ \bibinfo {author} {\bibfnamefont {Y.}~\bibnamefont {Chen}},\ }\bibfield  {title} {\bibinfo {title} {Nonperturbative phonon scatterings and the two-channel thermal transport in {Tl$_3$VSe$_4$}},\ }\href {https://doi.org/10.1103/physrevb.103.224307} {\bibfield  {journal} {\bibinfo  {journal} {Phys. Rev. B}\ }\textbf {\bibinfo {volume} {103}},\ \bibinfo {pages} {224307} (\bibinfo {year} {2021})}\BibitemShut {NoStop}%
\bibitem [{\citenamefont {Nataf}\ \emph {et~al.}(2024)\citenamefont {Nataf}, \citenamefont {Volz}, \citenamefont {Ordonez-Miranda}, \citenamefont {Íñiguez González}, \citenamefont {Rurali},\ and\ \citenamefont {Dkhil}}]{Nataf24p530}%
  \BibitemOpen
  \bibfield  {author} {\bibinfo {author} {\bibfnamefont {G.~F.}\ \bibnamefont {Nataf}}, \bibinfo {author} {\bibfnamefont {S.}~\bibnamefont {Volz}}, \bibinfo {author} {\bibfnamefont {J.}~\bibnamefont {Ordonez-Miranda}}, \bibinfo {author} {\bibfnamefont {J.}~\bibnamefont {Íñiguez González}}, \bibinfo {author} {\bibfnamefont {R.}~\bibnamefont {Rurali}},\ and\ \bibinfo {author} {\bibfnamefont {B.}~\bibnamefont {Dkhil}},\ }\bibfield  {title} {\bibinfo {title} {Using oxides to compute with heat},\ }\href {https://doi.org/10.1038/s41578-024-00690-1} {\bibfield  {journal} {\bibinfo  {journal} {Nat. Rev. Mater.}\ }\textbf {\bibinfo {volume} {9}},\ \bibinfo {pages} {530–531} (\bibinfo {year} {2024})}\BibitemShut {NoStop}%
\end{thebibliography}%
\end{document}